\newcommand{\eps}{\epsilon}
\newcommand{\nn}{\nonumber}
\newcommand{\ndotv}{({\bf n}'\cdot{\bf v})}
\newcommand{\ndotb}{({\bf n}'\cdot{\bf b})}
\newcommand{\w}{\omega}
\newcommand{\W}{\Omega}
\newcommand{\gkvec}[1]{{\mbox{\boldmath $#1$}}}
\newcommand{\dg}{\delta g}
\newcommand{\dgbar}{\overline{\delta g}}
\newcommand{\ttilde}{\tilde{t}}
\newcommand{\xt}{\tilde{x}}
\newcommand{\yt}{\tilde{y}}
\newcommand{\zt}{\tilde{z}}
\newcommand{\rt}{\tilde{r}}
\newcommand{\hatt}{\hat{t}}
\newcommand{\hx}{\hat{x}}
\newcommand{\hy}{\hat{y}}
\newcommand{\hz}{\hat{z}}
\newcommand{\hr}{\hat{r}}
\newcommand{\mh}{\hat{\mu}}
\newcommand{\nh}{\hat{\nu}}
\newcommand{\tc}{\check{t}}
\newcommand{\xc}{\check{x}}
\newcommand{\yc}{\check{y}}
\newcommand{\zc}{\check{z}}
\newcommand{\rc}{\check{r}}
\newcommand{\tb}{\bar{t}}
\newcommand{\xb}{\bar{x}}
\newcommand{\yb}{\bar{y}}
\newcommand{\zb}{\bar{z}}
\newcommand{\rb}{\bar{r}}
\newcommand{\Wb}{\bar{\W}}
\newcommand{\G}{\Gamma}
\newcommand{\Lm}{\Lambda}
\documentstyle[eqsecnum,aps,epsfig]{revtex}
\begin{document}
\draft

\title{An approximate binary-black-hole metric}
\author{Kashif Alvi}
\address{Theoretical Astrophysics, California Institute of Technology,
Pasadena, California 91125}
\maketitle
\begin{abstract}
An approximate solution to Einstein's equations representing two widely-separated
non-rotating black holes in a circular orbit is
constructed by matching a post-Newtonian metric to two perturbed Schwarzschild
metrics.  The spacetime metric is presented in a single coordinate system
valid up to the apparent horizons of the black holes.  This metric could be useful in
numerical simulations of binary black holes.  Initial data extracted from this metric have
the advantages of being linked to the early inspiral phase of the binary system, and of
not containing spurious gravitational waves.
\end{abstract}
\pacs{PACS numbers: 04.25.-g, 04.25.Nx, 04.30.Db, 04.70.-s}

\section{Introduction}

One of the outstanding issues in gravitational wave research is calculating
the wave output from the last stages of inspiral of binary black holes
(BBHs).  This intermediate binary black hole (IBBH) problem has been discussed
by Brady, Creighton, and Thorne \cite{bct}.  The purpose of this paper is to provide an
approximate four-dimensional BBH metric from which initial data can be extracted
and evolved numerically into and through the IBBH region.

The approach I take
is based on the work of Manasse \cite{manasse} and D'Eath
\cite{death1,death2}.  I consider two widely-separated non-rotating black
holes in a circular orbit.  The black holes' mass ratio is
not restricted---they can have comparable masses.  However, the masses are assumed
to be much smaller than the distance between them\footnote{Throughout this paper I use
geometrized units in which $G=c=1$.}.  As a result spacetime can be divided
into four regions, each with its own approximation scheme to solve Einstein's equations.
There is a strong-gravity region near each of the black holes which is
described by the Schwarzschild solution plus a perturbation due to the
companion's tidal field.  This perturbation is constrained
to satisfy the linearized Einstein equations (LEEs) about the Schwarzschild
metric.  The companion black hole's electric-type and
magnetic-type tidal fields are both taken into account
in calculating the perturbation.

Outside the strong-gravity
regions but within the near zone, the metric can be approximated by a
post-Newtonian expansion.  Further out is the radiation zone
which contains outgoing gravitational waves and can be described by a post-Minkowski
expansion of the metric.

There are overlap zones in this spacetime where the regions
described above intersect in pairs.  In the overlap zones, two different approximation
schemes---one from each of the two intersecting regions---are both valid.
The perturbative expansions produced by the two approximation
schemes are matched in the overlap zones using the framework of matched asymptotic
expansions.  The post-Newtonian near-zone metric---taken from \cite{bfp}---and
the radiation-zone metric---taken from
\cite{w&w}---already match in their overlap region.  In this paper,
the post-Newtonian near-zone metric is
matched to a perturbed Schwarzschild metric in the matching or buffer zone
surrounding each black hole.  This yields information on the asymptotic
behavior of
the Schwarzschild perturbation at large distances from the horizon, and on
the coordinate transformation between the two buffer-zone coordinate systems.

The Schwarzschild perturbation and coordinate transformation are not uniquely determined.
However, a different choice of
transformation---and hence different form of Schwarzschild perturbation---should
still represent the same physical situation.  In other words, different perturbations
that match to the post-Newtonian near-zone metric are expected to be related via gauge
transformations.  For the purposes of this paper, it is sufficient to find
one transformation and one Schwarzschild perturbation associated with each black hole
that result in a match between the post-Newtonian near-zone metric and
the distorted-black-hole metrics.

An approximate spacetime metric is put together by joining the regional
metrics at some specific 3-surfaces in the matching zones.  The final 4-metric is written
in a single coordinate system valid up to (but not inside) the apparent horizons of the black
holes.  This metric is useful not only as a source of initial data for
numerical evolution, but also as a check on the early stages of such an
evolution.

It has been
suggested that numerical simulation of BBHs should be performed in corotating
coordinates \cite{bct}.  For this reason the metric in final form
is given in corotating coordinates.   The BBH spacetime can be sliced and spatial coordinates
chosen in any convenient way when extracting initial data from the metric.  (Asymptotically
inertial coordinates can be used, for example.)

Initial data generated by the method presented in this paper have the advantage of being
connected to the early inspiral phase of the BBHs.  Detailed gravitational waveforms from
this early inspiral phase have already been calculated using post-Newtonian expansions.  These
waveforms will be easily linked to the waveforms obtained by evolving initial data extracted
from the metric presented here.

Initial data from this metric have the additional advantages of not containing
spurious gravitational waves and of reliably describing the physical situation of coalescing BBHs.
The accuracy of this description can be improved by taking the calculation in this paper
to higher orders. 

In Sec.~\ref{near&farzone}, the near-zone and radiation-zone metrics are written down.  In
Sec.~\ref{tidaldistortionBH1}, the first black hole's tidal deformation is calculated.  In
Sec.~\ref{finalformintmet}, the buffer-zone coordinate transformations are determined, and the
distorted-black-hole metrics are written in corotating post-Newtonian coordinates.  The
full spacetime metric is summarized in Sec.~\ref{results}.

\section{Near-zone and radiation-zone metrics}
\label{near&farzone}

Blanchet and collaborators (\cite{bdi} and references therein) and Will and Wiseman
\cite{w&w} have calculated in detail the near-zone and radiation-zone gravitational fields
of compact binary systems.  The approach taken by Will and Wiseman is particularly useful here
because they use a single coordinate system---harmonic coordinates---to cover
both the near zone and the
radiation zone.  As a result, expressions for the radiation-zone
metric components taken from \cite{w&w} automatically match (to some finite order) the
harmonic-coordinate, post-Newtonian, near-zone metric components calculated in \cite{bfp}.
For this reason I work initially in harmonic coordinates $(t',x',y',z')$ with the
origin of the spatial coordinates placed at the binary system's center of mass.
I use only the first post-Newtonian (1PN) metric, not the full 2.5PN metric given in
\cite{bfp}\footnote{Higher order versions of this calculation will presumably use higher order
post-Newtonian metrics.}.  Consistently with this, I put the black holes on Newtonian
trajectories: they are taken to be in circular orbits with Keplerian orbital angular
velocities.  Moreover, I use the post-Newtonian metric for point-like particles; in the
near zone, I ignore the black holes' internal structure.  The near-zone
gravitational effects of the black holes' multipole moments can in principle be
computed by matching out to the near zone the tidally-distorted Schwarzschild metrics
obtained in this paper.  However, these effects are too small to be included in this paper;
this is discussed further in Sec.~\ref{intcorotmet}.

\subsection{Binary-system parameters}

Label the black holes BH1 and BH2, and let $m_1$ and $m_2$ be their respective masses.
Define
\begin{equation}
	m = m_1 + m_2,\qquad \delta m = m_1 - m_2,\qquad \mu = {m_1 m_2\over m}.
\end{equation}
Denote the harmonic-coordinate trajectories of the black
holes by $x_A^j(t')$ for $A=1,2$ and $j=1,2,3$.  In other words, $x_A^j(t')$ are the
spatial coordinates at time $t'$ of the center
of attraction of the gravitational field of black hole $A$.

In this section, boldface letters
are used to denote spatial coordinates.  For example ${\bf x}_A=(x_A^1,
x_A^2,x_A^3)=(x_A,y_A,z_A)$.  The notation ${\bf a}\cdot{\bf b}$ is used for the quantity
$\delta_{jk} a^j b^k$, and $|{\bf a}|$ is by definition $({\bf a}\cdot{\bf a})^{1/2}$.

Denote the
black holes' separation $|{\bf x}_1-{\bf x}_2|$ by $b$.  The circular,
Newtonian trajectories of the black holes are
\begin{equation}
	{\bf x}_1(t') = \frac{m_2}{m} {\bf b}(t'),\qquad
		{\bf x}_2(t') = -{m_1\over m} {\bf b}(t')
\label{bhorbits}
\end{equation}
where
\begin{equation}
	{\bf b}(t') = {\bf x}_1(t')-{\bf x}_2(t') = b(\cos\w t',\sin\w t',0)
\label{b}
\end{equation}
and
\begin{equation}
	\w = \sqrt{{m\over b^3}}
\label{w}
\end{equation}
is the Keplerian orbital angular velocity.  Define
\begin{eqnarray}
	\eps &=& \sqrt{{m\over b}},\qquad r=(x'^2+y'^2+z'^2)^{1/2},\nn\\
	r_A &=& |{\bf x}'- {\bf x}_A|,\qquad
		{\bf n}_A = {{\bf x}'- {\bf x}_A\over r_A},\nn\\
	{\bf v}_A &=& {d{\bf x}_A\over dt'}, \qquad v_A=|{\bf v}_A|,\nn\\
	{\bf v} &=& {\bf v}_1 - {\bf v}_2 = \eps(-\sin\w t', \cos\w t', 0),\nn\\
\label{defns}
\end{eqnarray}
for $A=1,2$.  By assumption, $\eps\ll 1$.

\subsection{Demarcation of four regions in the BBH spacetime}

Let us first fix precisely four regions in this BBH spacetime; each of these regions will
receive a metric calculated as an approximate solution to the Einstein equations.  With such
a partition of spacetime in mind, define the inner limits
$r^{in}_1 = \sqrt{m_1 b}$ and $r^{in}_2 = \sqrt{m_2 b}$.
These are just convenient choices for the inner limits.  The important
property $r^{in}_1$ has is that both $r^{in}_1/b \to 0$ and $m_1/r^{in}_1
\to 0$ as $m_1/b \to 0$.  Similarly $r^{in}_2/b \to 0$ and $m_2/r^{in}_2
\to 0$ as $m_2/b \to 0$.  Also define
the outer limit $r^{out} = \lambda_c/2\pi = b/2\eps$ where $\lambda_c=\pi/\w$
is the characteristic wavelength of gravitational radiation emitted by
the binary system.

\begin{figure}
\begin{center}
\epsfig{file=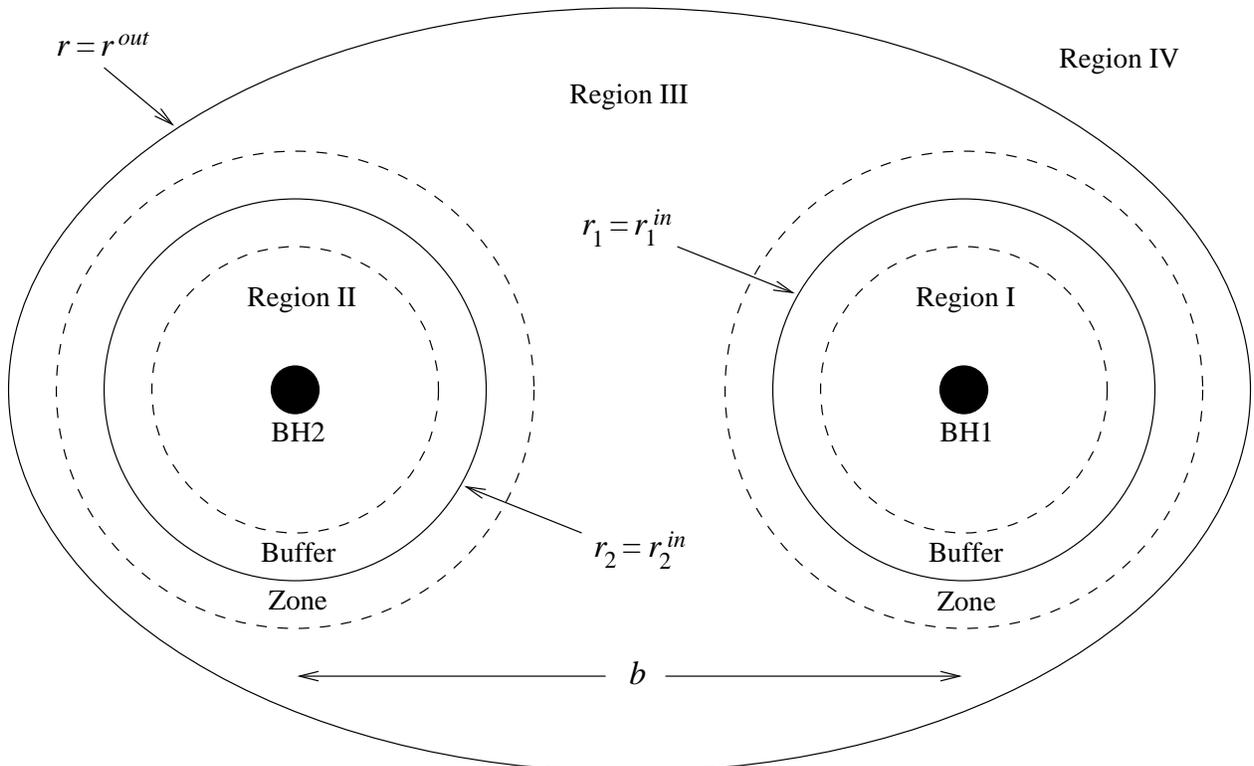}
\caption{\label{fig}
Schematic illustration of the various regions in the BBH spacetime.  Regions I, II, III,
and IV are demarcated by solid lines; the buffer zones are bounded by dashed lines.}
\end{center}
\end{figure}

Divide spacetime into four regions that are bounded by the black holes' apparent
horizons and the surfaces
$r_1 = r_1^{in}$, $r_2 = r_2^{in}$, and $r = r^{out}$: (i) the region
$r_1 < r_1^{in}$ (but outside the apparent horizon of BH1), labeled region I;
(ii) the region $r_2 < r_2^{in}$ (but outside the apparent horizon of BH2),
labeled region II; (iii) the subset of the near zone specified by
$r_1 > r_1^{in}$, $r_2 > r_2^{in}$, and $r < r^{out}$, labeled region III;
and (iv) the region $r > r^{out}$, labeled region IV.
The near zone contains
region III and overlaps with regions I and II; the radiation zone corresponds
to region IV.  The buffer zone around black hole A contains $r_A=r_A^{in}$ and
satisfies $m_A\ll r_A\ll b$.  These regions of spacetime are illustrated in
Fig.~\ref{fig}.

\subsection{Near-zone metric in harmonic coordinates}
\label{sec:harmonicmetnear}

In the near zone, the 1PN harmonic-coordinate metric with two
point-like particles representing the black holes is \cite{bfp}
\begin{eqnarray}
	g_{0'0'} &=& -1 + {2m_1 \over r_1} + {2m_2 \over r_2}
		- 2\left({m_1 \over r_1} + {m_2 \over r_2}\right)^2
		+ {m_1 \over r_1}\left[4v_1^2 - ({\bf n}_1\cdot{\bf v}_1)^2\right]
		+ {m_2 \over r_2}\left[4v_2^2 - ({\bf n}_2\cdot{\bf v}_2)^2\right]\nn\\
	& & \mbox{} - 2{m_1 m_2 \over b}\left({1\over r_1}+{1\over r_2}\right)
		+ {m_1 m_2 \over b^3}{\bf b}\cdot({\bf n}_1 - {\bf n}_2),\nn\\
	g_{0'i'} &=& -4\left({m_1 \over r_1} v_1^i + 
			{m_2 \over r_2} v_2^i \right),\nn\\
	g_{i'j'} &=& \delta_i{}_j \left(1 + {2m_1 \over r_1} +
			{2m_2 \over r_2} \right).
\label{1PN}
\end{eqnarray}
This metric presumably differs in the near zone by a small amount from an exact solution
to the Einstein equations representing BBHs.  I take the neglected terms in the 2.5PN
metric \cite{bfp} to be an estimate of the errors in the 1PN metric~(\ref{1PN}).

The largest neglected terms in $g_{0'0'}$ are of the form $m^3/b^2 r_A$, $m^3/br_A^2$,
$m_A^3/r_A^3$, $m^3/b^3$, and $\eps m^3 r^2/b^5$.  (The last term represents a radiation
reaction potential.)
Let us compute the orders of magnitude of these terms at various places in region~III.
If $r_A\gtrsim r_A^{in}\sim b\eps$ (here and henceforth ``$\sim$'' means
``is of the order of'' and $a\gtrsim b$ means $a>b$ and $a\sim b$)
for $A=1$ or 2, then the error in $g_{0'0'}$ (denoted
$\dg_{0'0'}$) is of $O(\eps^3)$ and comes from neglecting a term of the form $m_A^3/r_A^3$.
If both $r_1\sim b$ and $r_2\sim b$, then $\dg_{0'0'}\sim\eps^6$.  Finally, if
$r\lesssim r^{out}\sim b/\eps$ (so that $r_A\sim b/\eps$ for $A=1$ and 2), then the error
$\dg_{0'0'}\sim\eps^5$ arises from neglecting the radiation reaction potential.
Note that it is reasonable to consider the ``absolute'' errors $\dg_{\mu'\nu'}$
in the metric components since the coordinate system being used is asymptotically inertial
and the errors are only calculated in regions of weak gravity where deviations from a flat
metric are small.

A similar analysis for $g_{0'i'}$ yields $\dg_{0'i'}\sim\eps^3$ if
$r_A\gtrsim r_A^{in}$ for $A=1$ or 2, $\dg_{0'i'}\sim\eps^5$ if both $r_1\sim b$ and
$r_2\sim b$, and $\dg_{0'i'}\sim\eps^5$ if $r\lesssim r^{out}$.  Lastly,
$\dg_{i'j'}\sim\eps^2$ if
$r_A\gtrsim r_A^{in}$ for $A=1$ or 2 (this comes from neglecting a term of the form $m_A^2/r_A^2$
in $g_{i'j'}$), $\dg_{i'j'}\sim\eps^4$ if both $r_1\sim b$ and
$r_2\sim b$, and $\dg_{i'j'}\sim\eps^5$ if $r\lesssim r^{out}$.

\subsection{Near-zone metric in corotating coordinates}

The metric~(\ref{1PN}) is transformed to corotating coordinates $(t,x,y,z)$ defined by
\begin{eqnarray}
	t' &=& t,\qquad x' = x\cos\w t - y\sin\w t,\nn\\
	y' &=& x\sin\w t + y\cos\w t,\qquad z' = z.
\label{corotcoord}
\end{eqnarray}
In terms of the new coordinates,
\begin{equation}
	r = (x^2+y^2+z^2)^{1/2}.
\end{equation}
Putting the expressions~(\ref{bhorbits})--(\ref{defns}) in Eq.~(\ref{1PN}) and transforming
to corotating coordinates gives
\begin{eqnarray}
	ds^2 &=& dt^2 \biggl[-1 + {2m_1 \over r_1} + {2m_2 \over r_2}
		- 2\left({m_1 \over r_1} + {m_2 \over r_2}\right)^2
		+ {3\mu\over b}\left({m_2 \over r_1} + {m_1 \over r_2}\right)
		- {\mu\over b}\left({m_2 \over r_1^3} + {m_1 \over r_2^3}\right)y^2
		- 2\mu\eps^2\left({1\over r_1}+{1\over r_2}\right)\nn\\
	& & \mbox{} - 7\mu\eps^2\left({1\over r_1}-{1\over r_2}\right){x\over b}
		+ \w^2\left(1 + {2m_1 \over r_1} +
			{2m_2 \over r_2} \right)(x^2 + y^2)\biggr]
	     	+ 2\w\left(1 + {2m_1 \over r_1}
	     	+ {2m_2 \over r_2} \right)dt(x dy-y dx)\nn\\
	& & \mbox{} - 8\mu\eps\left({1\over r_1}
		- {1\over r_2}\right)dt dy + \left(1 + {2m_1 \over r_1} +
		{2m_2 \over r_2}\right)(dx^2+dy^2+dz^2),
\label{corotmetnear}
\end{eqnarray}
where, in terms of the new coordinates, the quantities $r_A$ are
\begin{eqnarray}
	r_1 &=& \left[(x-m_2 b/m)^2 + y^2 + z^2\right]^{1/2},\nn\\
	r_2 &=& \left[(x+m_1 b/m)^2 + y^2 + z^2\right]^{1/2}.
\end{eqnarray}

This is the final form of the metric in region III.  (Note, however, that this metric is
valid throughout the near zone, which includes the buffer zones around the black holes).
It remains to specify the metric in regions I, II, and IV.
I postpone until Sec.~\ref{sec:corotmetfar}
discussion of the errors $\dg_{\mu\nu}$ in the metric components in the new, rotating
coordinate system $(t,x,y,z)$.

\subsection{Radiation-zone metric in harmonic coordinates}
\label{sec:harmonicmetfar}

The radiation-zone metric can be extracted from \cite{w&w}.  In that paper, a potential
$h^{\mu'\nu'}$ is defined by
\begin{equation}
	h^{\mu'\nu'} = \eta^{\mu'\nu'}-(-g')^{1/2} g^{\mu'\nu'},
\label{hdef}
\end{equation}
where $\eta^{\mu'\nu'}$=diag(-1,1,1,1), $g_{\mu'\nu'}$ is the
spacetime metric, and $g'=\det(g_{\mu'\nu'})$.  Equation~(5.5) of \cite{w&w}
gives $h^{\mu'\nu'}$ in
the radiation zone in harmonic coordinates $(t',x',y',z')$ for a system of several bodies.
After correcting a typo in that equation\footnote{The
term $4m/r'$ in the expression for $h^{00}$ should instead be $4\tilde{m}/r'$.},
I specialize to a system of two bodies of masses $m_1$ and $m_2$ in a circular orbit
specified by Eqs.~(\ref{bhorbits})--(\ref{defns}).  This yields\footnote{Note that I
have replaced $r'$ in Eq.~(5.5) of \cite{w&w} by $r$.}
\begin{eqnarray}
	h^{0'0'}(t',x',y',z') &=& {4\tilde{m}\over r}+{7m^2\over r^2}+2\left[{1\over r}
		Q^{ij}(u')\right]_{,ij}-{2\over 3}\left[{1\over r}Q^{ijk}(u')\right]_{,ijk},\nn\\
	h^{0'i'}(t',x',y',z') &=& -2\left\{{1\over r}\left[\dot{Q}^{ij}(u')-\eps^{ijl}J^l (u')
		\right]\right\}_{,j}+{2\over 3}\left\{{1\over r}\left[\dot{Q}^{ijk}(u')
		-2\eps^{ikl}J^{lj}(u')\right]\right\}_{,jk},\nn\\
	h^{i'j'}(t',x',y',z') &=& {m^2\over r^2} n'{}^i n'{}^j+{2\over r}\ddot{Q}^{ij}(u')
		-{2\over 3}\left\{{1\over r}\left[\ddot{Q}^{ijk}(u')
		-4\eps^{(i|kl}\dot{J}^{l|j)}(u')\right]\right\}_{,k},
\label{h}
\end{eqnarray}
where
\begin{equation}
	\tilde{m} = m(1-\mu/2b),\qquad u' = t'-r,\qquad {\bf n}' = {\bf x}'/r
\end{equation}
and
\begin{eqnarray}
	Q^{ij} &=& \sum_{A=1}^{2} m_A x_A^i x_A^j = \mu b^i b^j,\qquad
	Q^{ijk} = \sum_{A=1}^{2} m_A x_A^i x_A^j x_A^k = -\mu(\delta m/m)b^i b^j b^k,\nn\\
	J^i &=& \sum_{A=1}^{2} m_A \eps^{ilm} x_A^l v_A^m = \mu \eps^{ilm} b^l v^m,\qquad
	J^{ij} = \sum_{A=1}^{2} m_A \eps^{ilm} x_A^l v_A^m x_A^j
		= -\mu (\delta m/m)\eps^{ilm} b^l v^m b^j.\nn\\
\label{q&j}
\end{eqnarray}
Putting the expressions~(\ref{q&j}) in Eq.~(\ref{h}) and using
Eqs.~(\ref{bhorbits})--(\ref{defns}) gives
\begin{eqnarray}
	h^{0'0'} &=& {4\tilde{m}\over r}+{7m^2\over r^2}+{2\mu\over r}\left\{2\ndotv^2
		-{2m\over b^3}\ndotb^2+{6\over r}\ndotb\ndotv
		+ {1\over r^2}\left[3\ndotb^2-b^2\right]\right\}\nn\\
	& & \mbox{}+{2\mu\over r}{\delta m\over m}\biggl\{{7m\over b^3}\ndotb^2\ndotv
		-2\ndotv^3+{1\over r}\ndotb\left[{6m\over b^3}\ndotb^2
		-12\ndotv^2-{m\over b}\right]\nn\\
	& & \mbox{}+{3\over r^2}\ndotv\left[b^2-5\ndotb^2\right]
		+{1\over r^3}\ndotb\left[3b^2-5\ndotb^2\right]\biggr\},\nn\\
	h^{0'i'} &=& {4\mu\over r}\left\{\left[\ndotv + {1\over r}\ndotb\right]v^i
		- {m\over b^3}\ndotb b^i\right\}-{2\mu\over r}{\delta m\over m}\biggl\{
		-{m\over b^3}\ndotb\left[3\ndotb v^i+4\ndotv b^i\right]\nn\\
	& & \mbox{}+2\ndotv^2 v^i+{1\over r}\left[6\ndotb\ndotv v^i
		-{3m\over b^3}\ndotb^2 b^i+{m\over b}b^i\right]+{1\over r^2}
		\left[3\ndotb^2-b^2\right]v^i\biggr\},\nn\\
	h^{i'j'} &=& {m^2\over r^2} n'{}^i n'{}^j + {4\mu\over r}\left[v^i v^j
		- {m\over b^3}b^i b^j\right]
		+{2\mu\over r}{\delta m\over m}\biggl\{{6m\over b^3}\ndotb v^{(i}b^{j)}\nn\\
	& & \mbox{}+\left[\ndotv+{1\over r}\ndotb\right]\left({m\over b^3} b^i b^j-2v^i v^j
		\right)\biggr\},
\label{htwo}
\end{eqnarray}
where ${\bf v}$ and ${\bf b}$ are evaluated at the retarded time $u'=t'-r$.

The metric $g_{\mu'\nu'}$ can be gotten from Eq.~(\ref{htwo}) as follows:
from Eq.~(\ref{hdef}) we have
\begin{equation}
	g^{\mu'\nu'} = (-g')^{-1/2}(\eta^{\mu'\nu'}-h^{\mu'\nu'}).
\label{gup}
\end{equation}
Take the determinant of both sides of Eq.~(\ref{gup}); this yields
$g'=\det(\eta^{\mu'\nu'}-h^{\mu'\nu'})$.  So $g'$ can be calculated
once $h^{\mu'\nu'}$ is known, and then $g^{\mu'\nu'}$ can be gotten from
Eq.~(\ref{gup}).  Inverting the matrix $g^{\mu'\nu'}$ gives the spacetime metric
$g_{\mu'\nu'}$.

When performing these calculations, I keep all terms of the form
$m^{3-p/2}b^{-3(1-p/2)}r^{-p}$ for integer $p>0$.  I also keep---at each order in $r$---all terms
that are of lower order in $m/b$ than this, and throw away terms of higher order in $m/b$.
This means in
particular that no terms of $O(r^{-5})$ are kept.  This
scheme of organizing terms is consistent with the ordering of terms in Eq.~(5.5) of \cite{w&w}.

The result of these calculations is
the following radiation-zone metric in harmonic coordinates:
\begin{eqnarray}
	g_{0'0'} &=& -1+{2\tilde{m}\over r}-{2m^2\over r^2}
		+{\mu\over r}\left\{2\ndotv^2-{2m\over b^3}\ndotb^2
		+{6\over r}\ndotb\ndotv + {1\over r^2}\left[3\ndotb^2-b^2\right]\right\}\nn\\
	& & \mbox{}+{\mu\over r}{\delta m\over m}\biggl\{\ndotv\left[{7m\over b^3}\ndotb^2
		-2\ndotv^2-{m\over b}\right]
		+{2\over r}\ndotb\left[{3m\over b^3}\ndotb^2-6\ndotv^2-{m\over b}\right]\nn\\
	& & \mbox{}+{3\over r^2}\ndotv\left[b^2-5\ndotb^2\right]
		+{1\over r^3}\ndotb\left[3b^2-5\ndotb^2\right]\biggr\},\nn\\
	g_{0'i'} &=& -{4\mu\over r}\left\{\left[\ndotv + {1\over r}\ndotb\right]v^i
		- {m\over b^3}\ndotb b^i\right\}+{2\mu\over r}{\delta m\over m}
		\Biggl(\biggl\{2\ndotv^2-{3m\over b^3}\ndotb^2\nn\\
	& & \mbox{}+{6\over r}\ndotb\ndotv+{1\over r^2}\left[3\ndotb^2-b^2\right]\biggr\}v^i
		+\left\{-{4m\over b^3}\ndotb\ndotv+{m\over rb}\left[1-{3\over b^2}
		\ndotb^2\right]\right\}b^i\Biggr),\nn\\
	g_{i'j'} &=& \delta_{ij}\Biggl(1+{2\tilde{m}\over r}+{m^2\over r^2}
		+{\mu\over r}\left\{2\ndotv^2
		-{2m\over b^3}\ndotb^2+{6\over r}\ndotb\ndotv+{1\over r^2}
		\left[3\ndotb^2-b^2\right]\right\}\nn\\
	& & \mbox{}+{\mu\over r}{\delta m\over m}\biggl\{\ndotv\left[{7m\over b^3}\ndotb^2
		-2\ndotv^2+{m\over b}\right]+{6\over r}\ndotb
		\left[{m\over b^3}\ndotb^2-2\ndotv^2\right]\nn\\
	& & \mbox{}+{3\over r^2}\ndotv\left[b^2-5\ndotb^2\right]+{1\over r^3}
		\ndotb\left[3b^2-5\ndotb^2\right]\biggr\}\Biggr)+{m^2\over r^2} n'{}^i n'{}^j
		+{4\mu\over r}\left(v^i v^j-{m\over b^3}b^i b^j\right)\nn\\
	& & \mbox{}+{2\mu\over r}{\delta m\over m}\left\{{6m\over b^3}\ndotb v^{(i}b^{j)}
		+\left[\ndotv+{1\over r}\ndotb\right]\left({m\over b^3}b^i b^j-2v^i v^j\right)
		\right\},
\label{farmet}
\end{eqnarray}
where ${\bf v}$ and ${\bf b}$ are evaluated at the retarded time
$u'=t'-r$.

The errors $\dg_{\mu'\nu'}$ in these metric components in region IV can be
estimated by computing the orders of magnitude of neglected terms, which are of the form 
$m^{3-p/2}b^{-3(1-p/2)}r^{-p}(m/b)^{n/2}$ for integers $p>0$ and $n>0$.  These terms include
$m^3/b^2 r$, $\eps m^2/r^2$, $m^2 b/r^3$, etc.  This gives $\dg_{\mu'\nu'}\sim\eps^7$
for $r\gtrsim r^{out}\sim b/\eps$, $\dg_{\mu'\nu'}\sim\eps^8$ for
$r\sim b/\eps^2\gg r^{out}$, and $\dg_{\mu'\nu'}\ll\eps^8$ for $r\gg b/\eps^2$.

\subsection{Radiation-zone metric in corotating coordinates}
\label{sec:corotmetfar}

Substituting the expressions~(\ref{bhorbits})--(\ref{defns}) into the metric~(\ref{farmet})
and transforming this metric to corotating coordinates $(t,x,y,z)$ [defined in
Eq.~(\ref{corotcoord})] gives
\begin{eqnarray}
	ds^2 &=& dt^2\biggl[-1 + {2m \over r}\left(1-{\mu\over 2b}\right)-{2m^2 \over r^2}+A
		+ {2\eps^2\over b}(x\cos\w r-y\sin\w r)B
		- 2\eps(x\sin\w r+y\cos\w r)D\nn\\
	& & \mbox{}+\w^2(x^2+y^2)E+{\eps^4\over b^2}(x\cos\w r-y\sin\w r)^2 N
		+\eps^2(x\sin\w r+y\cos\w r)^2S\nn\\
	& & \mbox{}-{12\mu\eps^5\over r^2 b^2}{\delta m\over m}(x\cos\w r-y\sin\w r)^2
		(x\sin\w r+y\cos\w r)\biggr]\nn\\
	& & \mbox{}+2dt\biggl(\eps[\sin(\w r)dx+\cos(\w r)dy]B
		+b[\cos(\w r)dx-\sin(\w r)dy]D+\w(xdy-ydx)E\nn\\
	& & \mbox{}+{\eps^3\over b}(x\cos\w r-y\sin\w r)
		[\sin(\w r)dx+\cos(\w r)dy]N\nn\\
	& & \mbox{}-\eps b(x\sin\w r+y\cos\w r)[\cos(\w r)dx-\sin(\w r)dy]S\nn\\
	& & \mbox{}+{6\mu\eps^4\over r^2 b}{\delta m\over m}(x\cos\w r-y\sin\w r)\left[
		(x\cos 2\w r-y\sin 2\w r)dx-(x\sin2\w r+y\cos2\w r)dy\right]\biggr)\nn\\
	& & \mbox{}+E(dx^2+dy^2+dz^2)+{m^2\over r^4}(xdx+ydy+zdz)^2+\eps^2
		[\sin(\w r)dx+\cos(\w r)dy]^2 N\nn\\
	& & \mbox{}+b^2[\cos(\w r)dx-\sin(\w r)dy]^2 S\nn\\
	& & \mbox{}+{12\mu\eps^3\over r^2}{\delta m\over m}(x\cos\w r-y\sin\w r)
		[\cos(\w r)dx-\sin(\w r)dy][\sin(\w r)dx+\cos(\w r)dy]
\label{corotmetfar}
\end{eqnarray}
where
\begin{eqnarray}
        A &=& {\mu\over r^3}\biggl\{2\eps^2\left[(x\sin\w r+y\cos\w r)^2
		-(x\cos\w r-y\sin\w r)^2\right]\nn\\
	& & \mbox{}+{6b\eps\over r}(x\sin\w r+y\cos\w r)(x\cos\w r-y\sin\w r)
		+{b^2\over r^2}\left[3(x\cos\w r-y\sin\w r)^2-r^2\right]\biggr\}\nn\\
        & & \mbox{}+{\mu\over r^4}{\delta m\over m}\biggl\{{b^3\over r^3}(x\cos\w r-y\sin\w r)
		\left[3r^2-5(x\cos\w r-y\sin\w r)^2\right]\nn\\
	& & \mbox{}+{3 b^2\eps\over r^2}
		(x\sin\w r+y\cos\w r)\left[r^2-5(x\cos\w r-y\sin\w r)^2\right]\nn\\
	& & \mbox{}+{2m\over r}(x\cos\w r-y\sin\w r)\left[3(x\cos\w r-y\sin\w r)^2
		-6(x\sin\w r+y\cos\w r)^2-r^2\right]\nn\\
	& & \mbox{}+\eps^3
		(x\sin\w r+y\cos\w r)\left[7(x\cos\w r-y\sin\w r)^2-2(x\sin\w r+y\cos\w r)^2
		-r^2\right]\biggr\},\nn\\
	B &=& -{4\mu\over r^2}\left[\eps(x\sin\w r+y\cos\w r)
		+{b\over r}(x\cos\w r-y\sin\w r)\right]\nn\\
	& & \mbox{}+{2\mu\over r^3}{\delta m\over m}\biggl\{\eps^2
		\left[2(x\sin\w r+y\cos\w r)^2-3(x\cos\w r-y\sin\w r)^2\right]\nn\\
	& & \mbox{}+{6b\eps\over r}(x\sin\w r+y\cos\w r)(x\cos\w r-y\sin\w r)
		+{b^2\over r^2}\left[3(x\cos\w r-y\sin\w r)^2-r^2\right]\biggr\},\nn\\
	D &=& {2\mu\eps^2\over r^2}\Biggl({2\over b}(x\cos\w r-y\sin\w r)
		-{1\over r}{\delta m\over m}
		\biggl\{{4\eps\over b}(x\cos\w r-y\sin\w r)(x\sin\w r+y\cos\w r)\nn\\
	& & \mbox{}+{1\over r}\left[3(x\cos\w r-y\sin\w r)^2-r^2\right]\biggr\}\Biggr),\nn\\
	E &=& 1 + {2m \over r}\left(1-{\mu\over 2b}\right)+{m^2\over r^2}+A
		+{2\mu\eps^2\over r^2}{\delta m\over m}\left[{b\over r}(x\cos\w r-y\sin\w r)
		+\eps(x\sin\w r+y\cos\w r)\right],\nn\\
	N &=& {4\mu\over r}\left\{1-{\delta m\over m}\left[{\eps\over r}(x\sin\w r+y\cos\w r)
		+{b\over r^2}(x\cos\w r-y\sin\w r)\right]\right\},\nn\\
	S &=& {2\mu\eps^2\over r b}\left\{-{2\over b}+{1\over r}{\delta m\over m}\left[
		{\eps\over b}(x\sin\w r+y\cos\w r)+{1\over r}(x\cos\w r-y\sin\w r)\right]\right\},
\label{ABDENS}
\end{eqnarray}
and $\eps=(m/b)^{1/2}$, $\w=(m/b^3)^{1/2}$.  This is the final form of the metric in region IV.

It is now necessary to evaluate the errors
$\dg_{\mu\nu}$ in the metric components in the corotating coordinate system $(t,x,y,z)$.
Since this coordinate system is not asymptotically inertial, it no longer makes sense
to compute absolute errors.  The rotation of the coordinates introduces terms of $O(\w r)$
and $O[(\w r)^2]$ in $g_{00}$ and terms of $O(\w r)$ in $g_{0i}$.  For this reason, I define
the ``normalized'' errors $\dgbar_{00}=\dg_{00}/(\w r)^2$, $\dgbar_{0i}=\dg_{0i}/\w r$,
and $\dgbar_{ij}=\dg_{ij}$ in region IV.  It follows that $\dgbar_{\mu\nu}\sim\eps^7$ for
$r\gtrsim r^{out}\sim b/\eps$, $\dgbar_{\mu\nu}\sim\eps^8$ for $r\sim b/\eps^2\gg r^{out}$,
and $\dgbar_{\mu\nu}\ll\eps^8$ for $r\gg b/\eps^2$.

In region III, $\w r$ is less than 1 (and in the buffer zones, $\w r\ll 1$).  So rotation of the
coordinates is not important in analyzing errors in the metric~(\ref{corotmetnear}) in
region III.  I continue to use absolute errors in that region.  The errors
$\dg_{\mu\nu}$ in the metric components~(\ref{corotmetnear}) in corotating coordinates
in region III are the same as the errors in harmonic coordinates
(see Sec.~\ref{sec:harmonicmetnear}):
(i) $\dg_{00}\sim\eps^3$, $\dg_{0i}\sim\eps^3$, and $\dg_{ij}\sim\eps^2$
if $r_A\gtrsim r_A^{in}$ for $A=1$ or 2; (ii) $\dg_{00}\sim\eps^6$, $\dg_{0i}\sim\eps^5$,
and $\dg_{ij}\sim\eps^4$ if both $r_1\sim b$ and $r_2\sim b$; and (iii)
$\dg_{\mu\nu}\sim\eps^5$ if $r\lesssim r^{out}$.

Since the analysis by Will and Wiseman \cite{w&w} of compact binary systems
uses a single coordinate chart to cover both the near and radiation zones, the
near-zone metric~(\ref{corotmetnear}) automatically matches (to some finite order; see below)
the radiation-zone metric~(\ref{corotmetfar}) at $r=r^{out}$.  The match is not perfect
because I have truncated the relevant perturbative expansions at finite order.  As a result,
there are discontinuities in the metric components at $r=r^{out}$.  The orders of magnitude of
these discontinuities can be estimated as follows: first expand $r_A^{-1}$ in powers of $b/r$
for $r>b$ and substitute this expansion in Eq.~(\ref{corotmetnear}); then expand
Eq.~(\ref{corotmetfar}) in powers of $\w r$ for $r<r^{out}$; finally, compare the two.  The
result is that the discontinuities in $g_{\mu\nu}$, denoted $[g_{\mu\nu}]$, are
$[g_{\mu\nu}]\sim\eps^5$ at $r=r^{out}$.

\section{Tidal deformation of the first black hole}
\label{tidaldistortionBH1}

The metric~(\ref{corotmetnear}) is valid not only in region III but also in the buffer
zones around the black holes.  The next step is to match
this metric to a tidally-distorted black-hole metric
in the buffer zone around BH1.
There are two coordinate systems which overlap in the buffer zone.  The first
is the corotating post-Newtonian coordinate system $(t,x,y,z)$ defined in
Eq.~(\ref{corotcoord}).  The second---to be called the internal coordinate
system---covers the strong-gravity region near
the first black hole and is valid from the black hole's apparent horizon up into
(and through) the buffer zone.  The internal coordinates are
chosen to be isotropic coordinates $(T,X,Y,Z)$ in which the unperturbed
Schwarzschild metric is
\begin{equation}
	ds^2 = -\left( {1-m_1/2R \over 1+m_1/2R}\right)^2 dT^2 +
		\left(1+{m_1\over 2R}\right)^4(dX^2+dY^2+dZ^2),
\label{isotropic}
\end{equation}
where
\begin{equation}
	R = (X^2+Y^2+Z^2)^{1/2}.
\label{R}
\end{equation}
The region these coordinates cover will be called the internal region; it contains region I
in particular.

The first step in matching the near-zone metric~(\ref{corotmetnear}) to the internal metric
[which is the Schwarzschild metric~(\ref{isotropic}) plus tidal perturbations] is to write the
metric~(\ref{corotmetnear}) in internal coordinates.  Then the near-zone metric and the
internal metric are both expanded in positive powers of $m_1/R$ and $R/b$ in the
buffer zone.  Finally, corresponding terms in the two asymptotic expansions are
equated.  The near-zone metric determines in this way the
asymptotic form of the tidal perturbations on BH1.  These perturbations are further
constrained to solve the LEEs about the
Schwarzschild metric~(\ref{isotropic}) and to be finite at the horizon
$R=m_1/2$.

The asymptotic form of the Schwarzschild perturbations in internal coordinates can be
determined---independently of the matching procedure described above---by calculating
the electric- and magnetic-type
tidal fields of BH2 in the buffer zone surrounding BH1.  Once this asymptotic form is known,
the matching procedure can be used to constrain the coordinate
transformation taking corotating coordinates $(t,x,y,z)$ to internal coordinates
$(T,X,Y,Z)$.  This is the approach to matching taken in this paper.  In the next two sections,
I calculate the second black hole's tidal fields and the perturbations they induce on the
first black hole.

\subsection{Tidal fields of the companion black hole}
\label{tidalfields}

Thorne and Hartle \cite{t&h} have analyzed the motion of an isolated black hole in an
arbitrary surrounding spacetime.  They define and discuss the black hole's local asymptotic
rest frame (LARF).  In the LARF of BH1, the metric can be expanded in powers of the black hole's
mass $m_1$ as follows\footnote{This expansion is written in Eq.~(2.5) of \cite{t&h}; I
have substituted $m_1$ for $M$ in that equation.}:
\begin{equation}
	g = g^{(0)}+m_1 g^{(1)}+m_1^2 g^{(2)}+\cdots.
\end{equation}
Here the metric $g^{(0)}$ represents the external universe without BH1; the
rest of the terms represent the black hole's internal gravitational field and the nonlinear
interaction between internal and external fields.  In this section, I will focus on the external
metric $g^{(0)}$ and use it to constrain internal perturbations.  Throughout this section
and in the rest of this paper,
boldface letters denote spacetime tensors of all ranks (including 4-vectors).

In the case of BBHs, the external metric is simply that of a single black hole; it is
the metric of the companion black hole BH2 of mass $m_2$.  This metric
must be expressed in LARF coordinates.  With this goal in mind, consider
first a freely-falling observer
in a circular, equatorial orbit around a Kerr black hole of mass $m_2$.  (I will later specialize
to a non-rotating black hole.)  The Kerr black hole represents BH2 while the observer's local
Lorentz frame (same as proper reference frame) represents the LARF of BH1.

The metric near the observer's world line is determined by the Kerr
black hole's electric-type and magnetic-type tidal fields as seen in the observer's local
Lorentz frame.  These
tidal fields can be evaluated by taking components of the
Kerr spacetime's Weyl tensor ${\bf C}$ in a
parallel-propagated orthonormal (PPON) tetrad along
the observer's world line.  The vectors in this tetrad form the
coordinate basis of the local Lorentz frame at the location of the
geodesic orbit.

The electric-type tidal field as seen by such an observer has been calculated
by Fishbone \cite{fishbone} and Marck \cite{marck}.  Marck
has computed a PPON tetrad $(\gkvec{\lambda}_0,\gkvec{\lambda}_1,\gkvec{\lambda}_2,
\gkvec{\lambda}_3)$ along arbitrary geodesics of the Kerr spacetime,
with $\gkvec{\lambda}_0$ equal to the 4-velocity
of the geodesic.  He obtains the electric-type tidal field by evaluating
\begin{equation}
	R_{0i0j} = {\bf C}(\gkvec{\lambda}_0,\gkvec{\lambda}_i,\gkvec{\lambda}_0,\gkvec{\lambda}_j).
\label{electidal}
\end{equation}

I specialize his tetrad to circular, equatorial geodesics; I also label the tetrad vectors
(and hence coordinate axes) differently.  Initially (that is, at proper time ${\cal T}=0$),
I choose $\gkvec{\lambda}_1$ to be radially outward (in Boyer-Lindquist coordinates);
$\gkvec{\lambda}_2$ is chosen so that the projections of $\gkvec{\lambda}_0$ and
$\gkvec{\lambda}_2$ on a constant-Boyer-Lindquist-time-$t$ surface are parallel; and
$\gkvec{\lambda}_3$ is then chosen to give
$(\gkvec{\lambda}_1,\gkvec{\lambda}_2,\gkvec{\lambda}_3)$ positive (i.e., right-handed)
orientation.  With this choice of tetrad, I obtain the magnetic-type tidal field using Marck's
work by evaluating
\begin{equation}
	R_{0ijk} = {\bf C}(\gkvec{\lambda}_0,\gkvec{\lambda}_i,\gkvec{\lambda}_j,\gkvec{\lambda}_k).
\label{magtidal}
\end{equation}
The results of the calculations~(\ref{electidal}) and (\ref{magtidal}) with the above choice
of tetrad are
\begin{eqnarray}
	R_{0101} &=& {m_2\over d^3}\left[1-3\left(1+{{\cal W}^2\over d^2}\right)
			\cos^2\Wb{\cal T}\right],\qquad
	R_{0202} = {m_2\over d^3}\left[1-3\left(1+{{\cal W}^2\over d^2}\right)
			\sin^2\Wb{\cal T}\right],\nn\\
	R_{0303} &=& {m_2\over d^3}\left(1+{3{\cal W}^2\over d^2}\right),\qquad
	R_{0102} = R_{0201} = -{3m_2\over d^3}\left(1+{{\cal W}^2\over d^2}\right)
			\cos\Wb{\cal T}\sin\Wb{\cal T},\nn\\
	R_{0112} &=& -R_{0121} = R_{0323} = -R_{0332} = {3m_2 {\cal W}\over d^4}\left(
			1+{{\cal W}^2\over d^2}\right)^{1/2}\cos\Wb{\cal T},\nn\\
	R_{0212} &=& -R_{0221} = R_{0331} = -R_{0313} = {3m_2 {\cal W}\over d^4}\left(
			1+{{\cal W}^2\over d^2}\right)^{1/2}\sin\Wb{\cal T},
\label{riemann}
\end{eqnarray}
and the rest of the Weyl-tensor components are zero.  Here ${\cal T}$ is proper time along
the geodesic, $m_2$ is the mass of the Kerr black
hole, $d$ is the Boyer-Lindquist radial coordinate of the circular, equatorial orbit, and
\begin{equation}
	\Wb = \sqrt{{m_2\over d^3}}
\end{equation}
is the (exact) rotation rate of the black hole's tidal field as seen in the local Lorentz frame.
The quantity ${\cal W}$ is given by
\begin{equation}
	{{\cal W}\over d} = \frac{\sqrt{m_2/d}\pm a/d}{\left(1-3m_2/d
		\mp 2a\sqrt{m_2/d^3}\right)^{1/2}},
\label{W}
\end{equation}
where $m_2 a$ is the black hole's angular momentum.
The upper sign in Eq.~(\ref{W}) is for a retrograde orbit while the lower one is for
a direct or prograde orbit.

Notice that for $d\gg m_2$, the electric- and magnetic-type tidal field components at
${\cal T}=0$ are simply related via a Lorentz boost with low velocity $(m_2/d)^{1/2}$.
For example,
\begin{equation}
	R_{0112}\bigl|_{{\cal T}=0} = -(m_2/d)^{1/2}\bigl[R_{0101}+R_{1212}
		\bigr]_{{\cal T}=0} = -(m_2/d)^{1/2}\bigl[2R_{0101}+R_{0202}\bigr]_{{\cal T}=0}
\end{equation}
to lowest order in $m_2/d$.  This fact will be used later in this section.

In the local Lorentz frame, the spacetime metric can be written as an expansion in powers of
distance ${\cal R}$ from the observer's geodesic world line \cite{m&m}.
The two types of tidal field~(\ref{electidal}) and (\ref{magtidal}) determine the metric
up to and including terms of $O({\cal R}^2)$.
After exploiting some gauge freedom (see Sec.~V.A.2 of \cite{membrane}
and Eq.~(2.7) of \cite{t&h}), the metric can be written in local coordinates $({\cal T},
{\cal X},{\cal Y},{\cal Z})$ as
\begin{eqnarray}
	g_{00} &=& -1 - R_{0i0j}({\cal T}) {\cal X}^i {\cal X}^j + O({\cal R}^3),\nn\\
	g_{0i} &=& -{2\over 3}R_{0jik}({\cal T}) {\cal X}^j {\cal X}^k + O({\cal R}^3),\nn\\
	g_{ij} &=& \delta_{ij}\left[1 - R_{0k0m}({\cal T}) {\cal X}^k {\cal X}^m\right]
		+ O({\cal R}^3),
\label{localmet}
\end{eqnarray}
where ${\cal R}=({\cal X}^2+{\cal Y}^2+{\cal Z}^2)^{1/2}$.  Substituting the
expressions~(\ref{riemann}) in Eq.~(\ref{localmet}) gives
\begin{eqnarray}
	g_{00} &=& -1+{m_2\over d^3}\left[3\left(1+{{\cal W}^2\over d^2}\right)
		({\cal X}\cos\Wb{\cal T}+{\cal Y}\sin\Wb{\cal T})^2-{\cal R}^2
		-{3{\cal W}^2\over d^2}{\cal Z}^2\right],\nn\\
	g_{0{\cal X}} &=& {2m_2{\cal W}\over d^4}\left(1+{{\cal W}^2\over d^2}\right)^{1/2}\left[
		({\cal Z}^2-{\cal Y}^2)\sin\Wb{\cal T}-{\cal X}{\cal Y}\cos\Wb{\cal T}\right],\nn\\
	g_{0{\cal Y}} &=& {2m_2{\cal W}\over d^4}\left(1+{{\cal W}^2\over d^2}\right)^{1/2}\left[
		({\cal X}^2-{\cal Z}^2)\cos\Wb{\cal T}+{\cal X}{\cal Y}\sin\Wb{\cal T}\right],\nn\\
	g_{0{\cal Z}} &=& {2m_2{\cal W}\over d^4}\left(1+{{\cal W}^2\over d^2}\right)^{1/2}
		({\cal Y}\cos\Wb{\cal T} - {\cal X}\sin\Wb{\cal T}){\cal Z},\nn\\
	g_{ij} &=& \delta_{ij}\left\{1+{m_2\over d^3}\left[3\left(1+{{\cal W}^2\over d^2}\right)
		({\cal X}\cos\Wb{\cal T}+{\cal Y}\sin\Wb{\cal T})^2-{\cal R}^2
		-{3{\cal W}^2\over d^2}{\cal Z}^2\right]\right\}
\label{tidalmet}
\end{eqnarray}
up to and including terms of $O({\cal R}^2)$.

The rotation rate $\Wb$ is only correct for test-particle orbits and is
exact in that case.  The correct rotation rate $\W$ of the second black hole's tidal
field---measured in a local inertial frame in the first black hole's LARF---is
actually determined by the post-Newtonian metric~(\ref{corotmetnear}) and by the
requirements that (i) this metric match the LARF metric [given in Eq.~(\ref{larfmet}) in
Sec.~\ref{pert} below]; and (ii) the LARF coordinate system be non-rotating relative to
local inertial frames.
The rotation rate $\W$ is calculated in Sec.~\ref{transform} by transforming the metric
(\ref{corotmetnear}) to internal coordinates and requiring a match to the
LARF metric~(\ref{larfmet}).  There it will be seen that the rotation rate
is\footnote{This rotation rate can also be calculated
by looking at geodetic precession of parallel-propagated vectors in the LARF \cite{thorne}.}
\begin{equation}
	\W = \w\left[1-{\mu\over b} + O(\eps^3)\right].
\label{Omega}
\end{equation}
Note that post-Newtonian corrections to the orbital angular velocity $\w$ of $O(\eps^2\w)$
have not been included in this paper.
 
The metric~(\ref{tidalmet}) is valid for all radii $d$ which allow a circular, equatorial,
geodesic orbit.  To apply Eq.~(\ref{tidalmet}) to the situation of widely-separated
non-rotating BBHs, I specialize to a Schwarzschild black hole by setting $a=0$ and
take the limit of small $m_2/d$, keeping only lowest-order terms in $m_2/d$.  [In particular,
I replace ${\cal W}/d$ with $(m_2/d)^{1/2}$.]  I then
replace $d$ with $b$, $\Wb$ with $\W$, and local coordinates
$({\cal T},{\cal X},{\cal Y},{\cal Z})$ with internal coordinates $(T,X,Y,Z)$ [which are
described above Eq.~(\ref{isotropic})].  The result is
\begin{eqnarray}
	g_{00} &=& -1+{m_2\over b^3}\left[3(X\cos\W T+Y\sin\W T)^2-R^2\right],\nn\\
	g_{0X} &=& {2m_2\over b^3}\sqrt{{m_2\over b}}\left[(Z^2-Y^2)\sin\W T- XY\cos\W T\right],\nn\\
	g_{0Y} &=& {2m_2\over b^3}\sqrt{{m_2\over b}}\left[(X^2-Z^2)\cos\W T+ XY\sin\W T\right],\nn\\
	g_{0Z} &=& {2m_2\over b^3}\sqrt{{m_2\over b}}(Y\cos\W T - X\sin\W T)Z,\nn\\
	g_{ij} &=& \delta_{ij}\left\{1+{m_2\over b^3}\left[3(X\cos\W T
		+ Y\sin\W T)^2- R^2\right]\right\},
\label{tidmet}
\end{eqnarray}
where $R=(X^2+Y^2+Z^2)^{1/2}$ as defined in Eq.~(\ref{R}).

The metric~(\ref{tidmet}) is still not applicable to BBHs since the observer was taken to be
massless [there are no factors of $m_1$ in Eq.~(\ref{tidmet})].  This can be fixed easily.
As mentioned above, the factors of $(m_2/b)^{1/2}$ in $g_{0i}$ in Eq.~(\ref{tidmet}) arise
from a Lorentz boost with low velocity $(m_2/b)^{1/2}$.  But the correct (Newtonian) relative
velocity between the black holes is $\eps=[(m_1+m_2)/b]^{1/2}$.  So I replace the factors
of $(m_2/b)^{1/2}$ in Eq.~(\ref{tidmet}) by $(m/b)^{1/2}$.
The resulting metric includes the second black hole's tidal fields but does not include the
first black hole's gravitational field:
\begin{eqnarray}
	g_{00} &=& -1+{m_2\over b^3}\left[3(X\cos\W T+Y\sin\W T)^2-R^2\right],\nn\\
	g_{0X} &=& {2m_2\over b^3}\sqrt{{m\over b}}\left[(Z^2-Y^2)\sin\W T- XY\cos\W T\right],\nn\\
	g_{0Y} &=& {2m_2\over b^3}\sqrt{{m\over b}}\left[(X^2-Z^2)\cos\W T+ XY\sin\W T\right],\nn\\
	g_{0Z} &=& {2m_2\over b^3}\sqrt{{m\over b}}(Y\cos\W T - X\sin\W T)Z,\nn\\
	g_{ij} &=& \delta_{ij}\left\{1+{m_2\over b^3}\left[3(X\cos\W T
		+ Y\sin\W T)^2- R^2\right]\right\}.
\label{larftidmet}
\end{eqnarray}
In the buffer zone around BH1, this metric provides the asymptotic form of the
perturbation on BH1.

\subsection{Schwarzschild perturbation}
\label{pert}

The next stage is to solve the LEEs about the Schwarzschild metric for a
perturbation which is finite at the horizon $R=m_1/2$ and asymptotes to the form
(\ref{larftidmet}) as $R/m_1 \to\infty$.  For ease in dealing with the LEEs, I transform to
spherical, isotropic, internal coordinates $(T,R,\theta,\phi)$ by letting
\begin{equation}
	X = R\sin\theta\cos\phi,\qquad Y = R\sin\theta\sin\phi,\qquad Z = R\cos\theta.
\end{equation}
The unperturbed Schwarzschild metric in these coordinates is
\begin{equation}
	ds^2 = -\left({1-m_1/2R \over 1+m_1/2R}\right)^2 dT^2 +
		\left(1+{m_1\over 2R}\right)^4\left[dR^2+R^2(d\theta^2+\sin^2\theta\, d\phi^2)\right].
\label{sphisotropic}
\end{equation}
The metric~(\ref{larftidmet}) in these coordinates is
\begin{eqnarray}
	ds^2 &=& -dT^2+dR^2+R^2(d\theta^2+\sin^2\theta\, d\phi^2)\nn\\
	& & \mbox{}-4{m_2 R^3\over b^3}\sqrt{{m\over b}}dT\left[\cos\theta\sin(\phi-\W T)
		d\theta+\sin\theta\cos(2\theta)\cos(\phi-\W T)d\phi\right]\nn\\
	& & \mbox{} + {m_2 R^2\over b^3}\left[3\sin^2\theta\cos^2(\phi-\W T)-1\right]
		\left[dT^2+dR^2+R^2(d\theta^2+\sin^2\theta\, d\phi^2)\right].
\label{sphmet}
\end{eqnarray}

The linearity of the LEEs allows me to look
separately for solutions corresponding to the electric-type and
magnetic-type tidal fields.  First I look for a perturbation ${\bf h}_1 = {\bf g}
-{\bf g}_s$ (where ${\bf g}_s$ is the unperturbed Schwarzschild metric and ${\bf g}$ is the full
metric including the perturbation) of BH1 which corresponds to the electric-type tidal field of
BH2 and is of the form
\begin{equation}
	{\bf h}_1 = {m_2 R^2\over b^3}\left[3\sin^2\theta\cos^2(\phi-\W T)
			-1\right]\left[f_1(R)dT^2 + f_2(R)dR^2
			+ f_3(R)R^2(d\theta^2+\sin^2\theta\, d\phi^2)\right],
\label{h1}
\end{equation}
as suggested by Eq.~(\ref{sphmet}).  In this notation, $dT$, $dR$, $d\theta$, and $d\phi$ are
coordinate one-forms and $dT^2$ denotes the tensor product $dT\otimes dT$.
The functions $f_1$, $f_2$, and $f_3$ are to be determined by solving the LEEs with
the following boundary conditions: (i) $f_1(R)$, $f_2(R)$, and $f_3(R)$ are
required to approach $1$ as $R/m_1 \to\infty$ so that the perturbation~(\ref{h1}) matches the
electric-type tidal field in Eq.~(\ref{sphmet}); and (ii) ${\bf h}_1$ is required to be
finite at $R=m_1/2$.

Consider solving the LEEs order by order in $\eps=(m/b)^{1/2}$.  Time derivatives of the
components of ${\bf h}_1$ produce factors of $m_1\W\sim\eps^3$ in the LEEs and can thus
be neglected.  A solution for ${\bf h}_1$ can then be found
using the Regge-Wheeler formalism \cite{r&w} for analysis
of stationary Schwarzschild perturbations.  Regge and Wheeler decompose perturbations
into even- and odd-parity modes and analyze them in a particular gauge chosen to simplify
computations.  In their
classification ${\bf h}_1$ is a superposition of static\footnote{Time dependence
in Eq.~(\ref{h1}) is to be ignored, as explained above.} even-parity modes
with angular numbers $l=2$ and $m=-2,0,2$.  The general solution of the LEEs for
static even-parity modes
with $l\geq 2$ is well-known in Schwarzschild coordinates and is given in Sec.~IV of
\cite{hartle}, for example.  A particular solution with $l=2$ that is finite at the black hole's
horizon and contains an arbitrary multiplicative constant is easily obtained from the general
solution, and is given in Eqs.~(6.5) and~(6.7) of
\cite{death1}, for example\footnote{The notation in Sec.~VI of \cite{death1}
may be confusing: $R$ there denotes a dimensionless quantity obtained from the Schwarzschild radial
coordinate $r_s$ by $R=r_s/M$ where $M$---in my notation $m_1$---is the mass of the black hole
being perturbed.}.  After transforming this solution to isotropic coordinates, the
multiplicative constant is determined by imposing the boundary condition (i)
(given at the end of the previous paragraph).
This yields the following solution for the radial factors $f_1(R)$, $f_2(R)$,
and $f_3(R)$ in isotropic coordinates:
\begin{eqnarray}
	f_1(R) &=& \left(1-{m_1\over 2R}\right)^4,\nn\\
	f_2(R) &=& \left(1-{m_1\over 2R}\right)^2\left(1+{m_1\over 2R}\right)^6,\nn\\
	f_3(R) &=& \left(1+{m_1\over 2R}\right)^4\left[\left(1+{m_1\over 2R}\right)^4
			-{2m_1^2\over R^2}\right].
\label{f123}
\end{eqnarray}

Next I look for a perturbation ${\bf h}_2 = {\bf g}-{\bf g}_s$ of BH1 corresponding to the
magnetic-type tidal field of BH2 and of the form
\begin{equation}
	{\bf h}_2 = -{4m_2\over b^3}\sqrt{m\over b}R^3 F(R) dT\left[\cos\theta
		\sin(\phi-\W T)d\theta + \sin\theta\cos2\theta\cos(\phi-\W T)d\phi\right],
\label{h2}
\end{equation}
as suggested by Eq.~(\ref{sphmet}).  The function $F$ is to be determined by solving the LEEs
with the following boundary conditions: (i) $F(R)\to 1$ as $R/m_1 \to\infty$ so that the perturbation
(\ref{h2}) matches the magnetic-type tidal field in Eq.~(\ref{sphmet}); and (ii)
${\bf h}_2$ finite at $R=m_1/2$.  As was done for ${\bf h}_1$, time dependence is ignored in
${\bf h}_2$ since time derivatives produce higher-order terms.  In the Regge-Wheeler
classification, ${\bf h}_2$ is a superposition of stationary odd-parity modes with angular numbers
$l=2$ and $m=-1,1$.  The general solution of the LEEs for stationary odd-parity modes that are finite
at the horizon and have $l\geq 2$ is given in Schwarzschild coordinates in Eq.~(38) of \cite{r&w}.
This solution is only determined up to a multiplicative constant.  The particular case $l=2$ is
easily obtained from the general solution, and is given in Eq.~(6.10)
of \cite{death1}, for example\footnotemark[8].  After transforming
this solution to isotropic coordinates, the
multiplicative constant is determined by imposing the boundary condition (i) [given below
Eq.~(\ref{h2})].  This yields the following solution for the radial factor $F(R)$ in isotropic
coordinates:
\begin{equation}
	F(R) = \left(1-{m_1\over 2R}\right)^2\left(1+{m_1\over 2R}\right)^4.
\label{F}
\end{equation}

The metric in the internal region near BH1 is now complete.
It is given by the Schwarzschild metric~(\ref{sphisotropic}) plus the perturbations~(\ref{h1})
and (\ref{h2}) with radial factors given in Eqs.~(\ref{f123}) and (\ref{F}); in other words,
${\bf g}={\bf g}_s+{\bf h}_1+{\bf h}_2$.  In spherical isotropic coordinates
$(T,R,\theta,\phi)$, this internal metric is
\begin{eqnarray}
	ds^2 &=& -\left({1-m_1/2R \over 1+m_1/2R}\right)^2 dT^2+\left(1+{m_1\over 2R}\right)^4
		\left[dR^2+R^2(d\theta^2+\sin^2\theta\, d\phi^2)\right]\nn\\
	& & \mbox{}-{4m_2\over b^3}\sqrt{{m\over b}}\left(1-{m_1\over 2R}\right)^2
		\left(1+{m_1\over 2R}\right)^4 R^3 dT\left[\cos\theta\sin(\phi-\W T)
		d\theta+\sin\theta\cos(2\theta)\cos(\phi-\W T)d\phi\right]\nn\\
	& & \mbox{}+{m_2 R^2\over b^3}\left[3\sin^2\theta\cos^2(\phi-\W T)-1\right]
		\biggl\{\left(1-{m_1\over 2R}\right)^4 dT^2+\left(1-{m_1\over 2R}\right)^2
		\left(1+{m_1\over 2R}\right)^6 dR^2\nn\\
	& & \mbox{}+\left(1+{m_1\over 2R}\right)^4
		\left[\left(1+{m_1\over 2R}\right)^4-{2m_1^2\over R^2}\right]
		R^2(d\theta^2+\sin^2\theta\, d\phi^2)\biggr\}.
\label{sphintmet}
\end{eqnarray}
In isotropic coordinates $(T,X,Y,Z)$, this metric is
\begin{eqnarray}
	g_{00} &=& -\left(\frac{1-m_1/2R}{1+m_1/2R}\right)^2
		+{m_2\over b^3}\left(1-{m_1\over 2R}\right)^4
		\left[3(X\cos\W T+Y\sin\W T)^2-R^2\right],\nn\\
	g_{0X} &=& {2m_2\over b^3}\sqrt{{m\over b}} \left
		(1-{m_1\over 2R}\right)^2 \left(1+{m_1\over 2R}
		\right)^4\left[(Z^2-Y^2)\sin\W T - XY\cos\W T\right],\nn\\
	g_{0Y} &=& {2m_2\over b^3}\sqrt{{m\over b}} \left
		(1-{m_1\over 2R}\right)^2 \left(1+{m_1\over 2R}
		\right)^4\left[(X^2-Z^2)\cos\W T + XY\sin\W T\right],\nn\\
	g_{0Z} &=& {2m_2\over b^3}\sqrt{{m\over b}} \left
		(1-{m_1\over 2R}\right)^2 \left(1+{m_1\over 2R}
		\right)^4 (Y\cos\W T- X\sin\W T)Z,\nn\\
	g_{ij} &=& \left(1+{m_1\over 2R}\right)^4 \Biggl(\delta_{ij}
		+{m_2\over b^3}\left[3(X\cos\W T+Y\sin\W T)^2-R^2\right] \nn\\
	& & \times\left\{\left[\left(1+{m_1\over 2R}\right)^4-{2m_1^2\over R^2}\right]
		\delta_{ij} -{2m_1\over R}\left(1+{m_1^2\over4R^2}
		\right){X^i X^j\over R^2}\right\}\Biggr).
\label{intmet}
\end{eqnarray}

Expanding the components~(\ref{intmet}) in positive powers of $m_1/R$ and
$R/b$ in the buffer zone $m_1 \ll R \ll b$ and keeping only
lowest-order terms yields the LARF metric:
\begin{eqnarray}
	g_{00} &=& -1+{2m_1\over R} +{m_2\over b^3}\left[3(X\cos\W T+
		Y\sin\W T)^2-R^2 \right],\nn\\
	g_{0X} &=& {2m_2\over b^3}\sqrt{m\over b}\left[(Z^2-Y^2)\sin\W T
		- XY\cos\W T \right],\nn\\
	g_{0Y} &=& {2m_2\over b^3}\sqrt{m\over b}\left[(X^2-Z^2)\cos\W T
		+ XY\sin\W T \right],\nn\\
	g_{0Z} &=& {2m_2\over b^3}\sqrt{m\over b}\left(Y\cos\W T
		- X\sin\W T \right)Z,\nn\\
	g_{ij} &=& \delta_{ij}\left\{1+{2m_1\over R}+{m_2\over b^3}\left[3(X\cos
		\W T + Y\sin\W T)^2- R^2 \right]\right\}.
\label{larfmet}
\end{eqnarray}
This metric includes the first black hole's (weak) gravitational field as well as the
second black hole's tidal fields.

\section{Distorted-black-hole metrics in corotating coordinates}
\label{finalformintmet}

The post-Newtonian metric~(\ref{corotmetnear}), when expressed in internal
coordinates $(T,X,Y,Z)$ in the buffer zone around BH1, must take the form~(\ref{larfmet}).
The next step is to find explicitly
the coordinate transformation in the buffer zone taking corotating post-Newtonian
coordinates to these internal coordinates.  Applying the inverse of this transformation to the
internal metric~(\ref{intmet}) will put that metric in corotating
coordinates $(t,x,y,z)$.  An identical procedure will then be followed to obtain the metric
near BH2 in corotating coordinates.

\subsection{Buffer-zone coordinate transformation}
\label{transform}

In this section, a series of coordinate transformations are performed on the metric
(\ref{corotmetnear}) in the buffer zone of BH1 to bring it to the form~(\ref{larfmet}).
Composing these transformations gives the final transformation
from corotating to internal coordinates.  Throughout this process terms of
$O(m^2)$ are dropped; justification for this will be given at the end of the section.

Begin with the near-zone metric~(\ref{corotmetnear})
with terms of $O(m^2)$ removed.  Restrict attention to the buffer zone
$m_1 \ll r_1 \ll b$ since this is where the corotating coordinate system
and internal coordinate system overlap.  Center the coordinate grid on BH1 by shifting
the origin to $(x,y,z)=(m_2 b/m,0,0)$.  This is done by defining a new
coordinate
\begin{equation}
	\xi = x-{m_2 b\over m}.
\label{xi}
\end{equation}
Next expand the metric in powers of the distance
\begin{equation}
	r_1 = (\xi^2+y^2+z^2)^{1/2}
\end{equation}
to the new origin.  The expansion for $r_2^{-1}$ is
\begin{equation}
	{1\over r_2} = \left[(b+\xi)^2+y^2+z^2\right]^{-1/2} = {1\over b} -
		{\xi\over b^2}+{2\xi^2-y^2-z^2 \over 2b^3} + \cdots.
\end{equation}
Positive powers of $r_1$ in the metric components
come in the form $(r_1/b)^p$ with integer $p>0$.  Since $r_1 \ll b$ in the buffer zone,
discard terms of $O[(r_1/b)^3]$ or higher.  This results in the following metric:
\begin{eqnarray}
	ds^2 &=& dt^2\left[-1+2m_1/r_1+(2m_2/b)(1+m_2/2m)+(m_2/b^3)(2\xi^2
		-y^2-z^2)+(m/b^3)(\xi^2+y^2)\right]\nn\\
	& & \mbox{}-2\w ydtd\xi\left[1+2m_2/b+2m_1/r_1
		-2m_2\xi/b^2\right]+2\w dtdy\Bigl\{(m_2/m)(b+4m_1+2m_2)-2\mu b/r_1 \nn\\
	& & \mbox{}+\xi(1+2m_1/r_1-2\mu/b)+(m_2/b^2)\left[2m_1\xi^2/m-(1+m_1/m)(y^2+z^2)
		\right]\Bigr\}\nn\\
	& & \mbox{}+(d\xi^2+dy^2+dz^2)\left[1+2m_1/r_1+2m_2/b-2m_2\xi/b^2+(m_2/b^3)(2\xi^2
		-y^2-z^2)\right].
\label{ximet}
\end{eqnarray}
Now renormalize the time-coordinate by defining
\begin{equation}
	t = \grave{t}\left[1+(m_2/b)(1+m_2/2m)\right],
\label{renormalize}
\end{equation}
and then perform a partial Lorentz transformation by setting
\begin{eqnarray}
	\grave{t} &=& \ttilde+(m_2\w/m)(b+4m_1+3m_2+m_2^2/2m)\tilde{y},\nn\\
	\xi &=& \xt,\qquad y = \yt,\qquad z = \zt.
\label{tilde}
\end{eqnarray}
In the new coordinates $(\ttilde,\xt,\yt,\zt)$, the metric~(\ref{ximet}) is
\begin{eqnarray}
	ds^2 &=& {d\ttilde}^2\left[-1+2m_1/\rt+(m_2/b^3)(3\xt^2-\rt^2)+(m/b^3)
		(\xt^2+\yt^2)\right]\nn\\
	& & \mbox{}-2\w d\ttilde d\xt\left\{\yt\left[1+(m_2/2mb)(6m_1+7m_2)
		+2m_1/\rt\right]-2m_2\xt\yt/b^2\right\}\nn\\
	& & \mbox{}+2\w d\ttilde d\yt\left\{\xt
		\left[1+(m_2/2mb)(-2m_1+3m_2)+2m_1/\rt\right]
		+(m_2/b^2)(3\xt^2-\yt^2-2\zt^2)\right\}\nn\\
	& & \mbox{}+(d\xt^2+d\yt^2+d\zt^2)\left[1+2m_1/\rt+2m_2/b-2m_2\xt/b^2
		+(m_2/b^3)(3\xt^2-\rt^2)\right]\nn\\
	& & \mbox{}+(m_2/b)d\yt\left[(m_2/m+2\xt/b) d\yt-(2\yt/b)d\xt\right]
\label{tildemet}
\end{eqnarray}
where $\rt=(\xt^2+\yt^2+\zt^2)^{1/2}$ and terms of $O(m^2)$ have been
dropped, as is done throughout this section.

Next clean up the spatial part of the metric by putting
\begin{eqnarray}
	\ttilde &=& \hatt,\qquad \xt = \hx(1-m_2/b)+(m_2/2b^2)(\hx^2+\hy^2-\hz^2),\nn\\
	\yt &=& \hy[1-(m_2/2mb)(2m_1+3m_2)],\qquad \zt=\hz(1-m_2/b)+(m_2/b^2)\hx\hz.
\label{hat}
\end{eqnarray}
Transforming the metric~(\ref{tildemet}) using Eq.~(\ref{hat}) results in
\begin{eqnarray}
	ds^2 &=& {d\hatt}^2\left[-1+2m_1/\hr+(m_2/b^3)(3\hx^2-\hr^2)+(m/b^3)
		(\hx^2+\hy^2)\right]-2\w d\hatt d\hx\left[\hy(1+m_2/b+2m_1/\hr)
		-m_2\hx\hy/b^2\right]\nn\\
	& & \mbox{}+2\w d\hatt d\hy\left\{\hx\left[1-(m_2/mb)(3m_1+m_2)+2m_1/\hr\right]
		+(m_2/2b^2)(7\hx^2-3\hy^2-5\hz^2)\right\}+2(m_2\w/b^2)\hy\hz d\hatt d\hz \nn\\
	& & \mbox{}+(d\hx^2+d\hy^2+d\hz^2)
		\left[1+2m_1/\hr+(m_2/b^3)(3\hx^2-\hr^2)\right]
\label{hatmet}
\end{eqnarray}
where $\hr=(\hx^2+\hy^2+\hz^2)^{1/2}$.

Focus attention on the terms $2\w d\hatt\left\{\hx d\hy\left[1-(m_2/mb)(3m_1+m_2)+2m_1/\hr
\right]-\hy d\hx(1+m_2/b+2m_1/\hr)\right\}$ in Eq.~(\ref{hatmet}).  These terms contain
information about the rotation of the
coordinate axes.  However, they are not yet in the form of the rotation terms
$2\W(1+2m_1/\hr)d\hatt(\hx d\hy-\hy d\hx)$ that result from rotating---at a constant rate $\W$
and in an active sense, i.e., using a pull-back map---the metric $ds^2={d\hatt}^2
(-1+2m_1/\hr)+(d\hx^2+d\hy^2+d\hz^2)(1+2m_1/\hr)$, which is a fragment of Eq.~(\ref{hatmet}).
An additional coordinate transformation
is required to bring the former terms into the latter form.  With this goal in mind,
look first for a gauge transformation taking the perturbation
\begin{equation}
	\gkvec{\gamma} = 2\w d\hatt\left\{\hx d\hy\left[1-(m_2/mb)(3m_1+m_2)+2m_1/\hr
		\right]-\hy d\hx(1+m_2/b+2m_1/\hr)\right\}
\label{gamma}
\end{equation}
on a flat background metric $ds^2=-d\hatt^2+d\hx^2+d\hy^2+d\hz^2$ to
the perturbation
\begin{equation}
	\breve{\gkvec{\gamma}} = 2\W(1+2m_1/\hr)d\hatt(\hx d\hy-\hy d\hx).
\label{newgamma}
\end{equation}
In other words, look for a vector field $\gkvec{\eta}$ such that
\begin{equation}
	\breve{\gamma}_{\mh\nh} = \gamma_{\mh\nh}-2\partial_{(\mh}\eta_{\nh)}.
\label{changegauge}
\end{equation}

In order to solve Eq.~(\ref{changegauge}), it suffices to consider $\gkvec{\eta}$ with only
one nonzero component
$\eta^{\hatt}=\eta^{\hatt}(\hx,\hy,\hz)$.  The perturbations~(\ref{gamma}) and
(\ref{newgamma}) when put in Eq.~(\ref{changegauge}) yield
\begin{eqnarray}
	-\W\hy(1+2m_1/\hr) &=& \breve{\gamma}_{\hatt\hx}=\gamma_{\hatt\hx}+\partial\eta^{\hatt}/
		\partial\hx=-\w\hy(1+m_2/b+2m_1/\hr)+\partial\eta^{\hatt}/\partial\hx,
\label{gammatx}\\
	\W\hx(1+2m_1/\hr) &=& \breve{\gamma}_{\hatt\hy}=\gamma_{\hatt\hy}+\partial\eta^{\hatt}/
		\partial\hy=\w\hx\left[1-(m_2/mb)(3m_1+m_2)+2m_1/\hr\right]
		+\partial\eta^{\hatt}/\partial\hy.
\label{gammaty}
\end{eqnarray}
These two equations determine the rotation rate $\W$ as follows: the function
$\eta^{\hatt}(\hx,\hy,\hz)$ must satisfy $\partial^2\eta^{\hatt}/\partial\hx\partial\hy=
\partial^2\eta^{\hatt}/\partial\hy\partial\hx$.
Taking $\partial/\partial\hy$ of Eq.~(\ref{gammatx}) and $\partial/\partial\hx$ of
Eq.~(\ref{gammaty}), equating the mixed partials of $\eta^{\hatt}$, and ignoring terms of
$O(m^2)$ yields the following equation for $\W$:
\begin{equation}
	\W-\w\left[1-(m_2/b)(1+2m_1/m)\right]=\w(1+m_2/b)-\W,
\label{eqnforW}
\end{equation}
which has solution $\W=\w\left[1-\mu/b + O(\eps^3)\right]$.  This is the rotation rate of
the second black hole's tidal field as seen in the first black hole's LARF; this value
confirms the claim in Sec.~\ref{tidalfields} [see Eq.~(\ref{Omega})].
With $\W$ in hand, Eqs.~(\ref{gammatx}) and (\ref{gammaty}) now yield
\begin{equation}
	\eta^{\hatt} = (m_2\w/b)(1+m_1/m)\hx\hy.
\label{eta}
\end{equation}

Gauge transformations can also be
thought of as resulting from infinitesimal coordinate transformations.
The coordinate transformation corresponding to the gauge transformation given in
Eqs.~(\ref{changegauge}) and (\ref{eta}) is
\begin{eqnarray}
	\hatt &=& \check{t}-(m_2\w/b)(1+m_1/m)\check{x}\check{y},\nn\\
	\hx &=& \check{x},\qquad \hy = \check{y},\qquad \hz = \check{z}.
\label{check}
\end{eqnarray}
The metric~(\ref{hatmet}) expressed in the new coordinates $(\tc,\xc,\yc,\zc)$ is
\begin{eqnarray}
	ds^2 &=& {d\tc}^2\left[-1+2m_1/\rc+(m_2/b^3)(3\xc^2-\rc^2)+(m/b^3)
		(\xc^2+\yc^2)\right]+2\W(1+2m_1/\rc)d\tc(\xc d\yc-\yc d\xc)\nn\\
	& & \mbox{}+(m_2\w/b^2)d\tc\left[2\xc\yc d\xc+(7\xc^2-3\yc^2-5\zc^2)d\yc
		+2\yc\zc d\zc\right]\nn\\
	& & \mbox{}+(d\xc^2+d\yc^2+d\zc^2)\left[1+2m_1/\rc+(m_2/b^3)(3\xc^2-\rc^2)\right],
\label{checkmet}
\end{eqnarray}
where $\rc=(\xc^2+\yc^2+\zc^2)^{1/2}$.

The next step is to undo the rotation of the coordinate system.  But first some fine-adjustment
of coordinates is needed in order to obtain the LARF metric~(\ref{larfmet}).
To find out what is
required, the metric~(\ref{larfmet}) can be put in coordinates rotating with angular
velocity $\W$.  It turns out that the fine-adjustment needed is
\begin{eqnarray}
	\check{t} &=& \tb+(m_2\w/2b^2)(3\xb^2-\yb^2-\zb^2)\yb,\nn\\
	\check{x} &=& \xb,\qquad \check{y} = \yb,\qquad \check{z} = \zb.
\label{bar}
\end{eqnarray}
In the new coordinates $(\tb,\xb,\yb,\zb)$, the metric~(\ref{checkmet}) is
\begin{eqnarray}
	ds^2 &=& {d\tb}^2\left[-1+2m_1/\rb+(m_2/b^3)(3\xb^2-\rb^2)+(m/b^3)
		(\xb^2+\yb^2)\right]+2\W(1+2m_1/\rb)d\tb(\xb d\yb-\yb d\xb)\nn\\
	& & \mbox{}+4(m_2\w/b^2)d\tb\left[-\xb\yb d\xb+(\xb^2-\zb^2)d\yb+\yb\zb d\zb\right]
		+(d\xb^2+d\yb^2+d\zb^2)\left[1+2m_1/\rb+(m_2/b^3)(3\xb^2-\rb^2)\right],
\label{barmet}
\end{eqnarray}
where $\rb=(\xb^2+\yb^2+\zb^2)^{1/2}$.
Now eliminate the rotation of coordinates by defining
\begin{eqnarray}
	\tb &=& T,\qquad \xb = X\cos\W T+Y\sin\W T,\nn\\
	\yb &=& -X\sin\W T+Y\cos\W T,\qquad \zb = Z.
\label{derotate}
\end{eqnarray}
Transforming the metric~(\ref{barmet}) using Eq.~(\ref{derotate}) results in the LARF metric
(\ref{larfmet}).

The transformation from corotating post-Newtonian coordinates $(t,x,y,z)$ to isotropic internal
coordinates $(T,X,Y,Z)$ can now be gotten by composing the transformations~(\ref{xi}),
(\ref{renormalize}), (\ref{tilde}), (\ref{hat}), (\ref{check}), (\ref{bar}), and
(\ref{derotate}).  Inverting this composite map gives the following
transformation from internal to corotating coordinates:
\begin{eqnarray}
	T &=& t\left[1-{m_2\over b}\left(1+{m_2\over 2m}\right)
		\right] - y\left[{m_2\over\sqrt{mb}} +
		{m_2\over b}\sqrt{{m\over b}}\left(3+{m_1\over m}
		+{m_2^2\over 2m^2}\right)\right] \nn\\
	  & & \mbox{}+{m_2 y\over b^2}\sqrt{{m\over b}}\left[\left(1+
		{m_1\over m}\right)\xi-\frac{1}{2b}\left(3\xi^2-y^2-z^2\right)\right],\nn\\
	X &=& \G\cos\W T - \Lm\sin\W T,\nn\\
	Y &=& \G\sin\W T + \Lm\cos\W T,\nn\\
	Z &=& z\left(1+{m_2\over b} -{m_2\xi\over b^2}\right),
\label{finaltrans}
\end{eqnarray}
where
\begin{eqnarray}
	\xi &=& x-{m_2b\over m},\qquad
	\G = \xi\left(1+{m_2\over b}\right)-{m_2\over 2b^2}(\xi^2+y^2-z^2),\nn\\
	\Lm &=& y\left[1+{m_2\over b}\left(1+{m_2\over 2m}\right)\right],\qquad
	\W = \w\left(1-{\mu\over b}\right),
\label{xGLW}
\end{eqnarray}
and terms of $O(m^2)$ have been dropped.  In terms of the coordinates $(x,y,z)$,
\begin{equation}
	R = (\G^2+\Lm^2+Z^2)^{1/2}
\label{newR}
\end{equation}
[cf. Eq.~(\ref{R})].

There are two reasons why terms of $O(m^2)$ were dropped from the metric~(\ref{corotmetnear})
at the beginning of this section.  First, suppose that such terms were kept and were used
to calculate higher-order deformation of the black hole.  Since internal metric components
are coupled to each other via the Einstein equations (in particular, the components of a black hole
perturbation are coupled via the LEEs), to be fully consistent, terms of $O(m^2)$ would
have to be included in the spatial part $g_{ij}$ of the metric~(\ref{corotmetnear}).
But these terms are of higher order than first post-Newtonian, and so have not been included
in this paper.

Second, black-hole perturbations with asymptotic form $m^2 r_1^p/b^{p+2}$ ($p\geq 2$) in the buffer
zone, which come from terms of $O(m^2)$ in $g_{00}$ in Eq.~(\ref{corotmetnear}), are actually
smaller in the internal region than the perturbation with asymptotic form $m_2 r_1^3/b^4$ in the
buffer zone; the latter perturbation has been ignored in this paper.  Once terms of $O(m^2)$ were
dropped in Eq.~(\ref{corotmetnear}), all terms of $O(m^2)$ were consistently discarded in this
section.

\subsection{Internal metric in corotating coordinates}
\label{intcorotmet}

In this section, the transformation~(\ref{finaltrans}) is applied to the internal metric
(\ref{intmet}) throughout region I (not just in the buffer zone).  This puts the internal
metric in corotating post-Newtonian coordinates $(t,x,y,z)$.
In order to preserve finiteness of the perturbations
(\ref{h1})--(\ref{f123}) and~(\ref{h2})--(\ref{F}) at the horizon of BH1, all terms must
be kept when performing the transformation.  The rotation in Eq.~(\ref{finaltrans}) can
easily be performed on the metric~(\ref{sphintmet}) by first defining $\varphi=\phi-\W T$ and
then setting $\G=R\sin\theta\cos\varphi$, $\Lm=R\sin\theta\sin\varphi$, and $Z=R\cos\theta$.
To complete the
transformation~(\ref{finaltrans}), define the functions $P_{\alpha\beta}(x,y,z)$ for
$\alpha,\beta=0,..,3$ to be components of the internal metric in
coordinates $(T,\G,\Lm,Z)$; write the components
as functions of $(x,y,z)$ using Eqs.~(\ref{finaltrans}) and (\ref{xGLW}).
Explicitly, the functions $P_{\alpha\beta}$ are
\begin{eqnarray}
P_{00} &=& -\left(\frac{1-m_1/2R}{1+m_1/2R}\right)^2
	+{m_2\over b^3}\left(1-{m_1\over 2R}\right)^4 (3\G^2-R^2)
	-{4\eps m_2\over b^3}\W\G\left(1-{m_1\over 2R}\right)^2
	\left(1+{m_1\over 2R}\right)^4(2Z^2-R^2)\nn\\
& &\mbox{}+\W^2\left(1+{m_1\over 2R}\right)^4(\G^2+\Lm^2)\left\{1+{m_2\over b^3}(3\G^2-R^2)
	\left[\left(1+{m_1\over 2R}\right)^4-{2m_1^2\over R^2}\right]\right\},\nn\\
P_{01} &=& P_{10} = -{2\eps m_2\over b^3}\left(1-{m_1\over 2R}\right)^2 \left(1+{m_1\over 2R}
	\right)^4\G\Lm\nn\\
& &\quad\qquad\mbox{}-\W\Lm\left(1+{m_1\over 2R}\right)^4\left\{1+{m_2\over b^3}(3\G^2-R^2)
	\left[\left(1+{m_1\over 2R}\right)^4-{2m_1^2\over R^2}\right]\right\},\nn\\
P_{02} &=& P_{20} = {2\eps m_2\over b^3}\left(1-{m_1\over 2R}\right)^2 \left(1+{m_1\over 2R}
	\right)^4(\G^2-Z^2)\nn\\
& &\quad\qquad\mbox{}+\W\G\left(1+{m_1\over 2R}\right)^4\left\{1+{m_2\over b^3}(3\G^2-R^2)
	\left[\left(1+{m_1\over 2R}\right)^4-{2m_1^2\over R^2}\right]\right\},\nn\\
P_{03} &=& P_{30} = {2\eps m_2\over b^3}\left(1-{m_1\over 2R}\right)^2
	\left(1+{m_1\over 2R}\right)^4 \Lm Z,\nn\\
P_{11} &=& \left(1+{m_1\over 2R}\right)^4 \left\{1+{m_2\over b^3}(3\G^2-R^2)
	\left[\left(1+{m_1\over 2R}\right)^4-{2m_1^2\over R^2}-{2m_1\over R^3}
	\left(1+{m_1^2\over4R^2}\right)\G^2\right]\right\},\nn\\
P_{22} &=& \left(1+{m_1\over 2R}\right)^4 \left\{1+{m_2\over b^3}(3\G^2-R^2)
	\left[\left(1+{m_1\over 2R}\right)^4-{2m_1^2\over R^2}-{2m_1\over R^3}
	\left(1+{m_1^2\over4R^2}\right)\Lm^2\right]\right\},\nn\\
P_{33} &=& \left(1+{m_1\over 2R}\right)^4 \left\{1+{m_2\over b^3}(3\G^2-R^2)
	\left[\left(1+{m_1\over 2R}\right)^4-{2m_1^2\over R^2}-{2m_1\over R^3}
	\left(1+{m_1^2\over4R^2}\right)Z^2\right]\right\},\nn\\
P_{12} &=& P_{21} = -{2m_1 m_2\over R^3 b^3}\left(1+{m_1\over 2R}\right)^4
	\left(1+{m_1^2\over4R^2}\right)(3\G^2-R^2)\G\Lm,\nn\\
P_{13} &=& P_{31} = -{2m_1 m_2\over R^3 b^3}\left(1+{m_1\over 2R}\right)^4
	\left(1+{m_1^2\over4R^2}\right)(3\G^2-R^2)\G Z,\nn\\
P_{23} &=& P_{32} = -{2m_1 m_2\over R^3 b^3}\left(1+{m_1\over 2R}\right)^4
	\left(1+{m_1^2\over4R^2}\right)(3\G^2-R^2)\Lm Z,
\label{P}
\end{eqnarray}
where $\eps=(m/b)^{1/2}$, $\W=\w(1-\mu/b)$, $R=(\G^2+\Lm^2+Z^2)^{1/2}$, and $\G$, $\Lm$,
and $Z$ are given in terms of $(x,y,z)$ in Eqs.~(\ref{finaltrans}) and~(\ref{xGLW}).

Next define the functions
$K_{\rho}^{\sigma}(x,y,z)$ for $\rho,\sigma=0,..,3$ by
$K^0_\rho=\partial T/\partial x^{\rho}$,
$K^1_\rho=\partial \G/\partial x^{\rho}$,
$K^2_\rho=\partial \Lm/\partial x^{\rho}$, and
$K^3_\rho=\partial Z/\partial x^{\rho}$, where $(T,\G,\Lm,Z)$ are to be expressed in
terms of $(t,x,y,z)$ using Eqs.~(\ref{finaltrans}) and (\ref{xGLW}).
Explicitly, the functions $K_{\rho}^{\sigma}$ are
\begin{eqnarray}
K^0_0 &=& 1-{m_2\over b}\left(1+{m_2\over 2m}\right),\qquad
K^0_1 = {\eps m_2 y\over b^2}\left(1+{m_1\over m}-{3\xi\over b}\right),\nn\\
K^0_2 &=& -\eps\left[{m_2\over m}+{m_2\over b}\left(3+{m_1\over m}
	+{m_2^2\over 2m^2}\right)\right]+{\eps m_2\over b^2}\left[\left(1+{m_1\over m}
	\right)\xi-{1\over 2b}\left(3\xi^2-3y^2-z^2\right)\right],\nn\\
K^0_3 &=& {\eps m_2\over b^3}yz,\nn\\
K^1_0 &=& 0,\qquad
K^1_1 = 1+{m_2\over b}-{m_2\xi\over b^2},\qquad
K^1_2 = -{m_2 y\over b^2},\qquad
K^1_3 = {m_2 z\over b^2},\nn\\
K^2_0 &=& 0,\qquad
K^2_1 = 0,\qquad
K^2_2 = 1+{m_2\over b}\left(1+{m_2\over 2m}\right),\qquad
K^2_3 = 0,\nn\\
K^3_0 &=& 0,\qquad
K^3_1 = -{m_2 z\over b^2},\qquad
K^3_2 = 0,\qquad
K^3_3 = 1+{m_2\over b}-{m_2\xi\over b^2},
\label{K}
\end{eqnarray}
where $\eps=(m/b)^{1/2}$ and $\xi=x-m_2 b/m$.

The metric in region I can now be written in terms of the functions $P_{\alpha\beta}$
and $K_{\rho}^{\sigma}$.  It is given in corotating coordinates $(t,x,y,z)$ by
\begin{equation}
	g_{\mu\nu}(x,y,z) = \sum_{\alpha,\sigma=0}^{3} P_{\alpha\sigma}(x,y,z)
		K^{\alpha}_{\mu}(x,y,z)K^{\sigma}_{\nu}(x,y,z)
\label{gPKI}
\end{equation}
with $P_{\alpha\beta}$ and $K_{\rho}^{\sigma}$ as defined in Eqs.~(\ref{P}) and~(\ref{K}).
Note that the metric components are explicitly independent of time $t$.
This metric is valid throughout region I (up to the first black hole's apparent horizon)
and matches (to some finite order; see below)
the post-Newtonian metric~(\ref{corotmetnear}) at $r_1=r_1^{in}$.

Errors in the internal metric~(\ref{gPKI}) will only be analyzed in the weak-gravity buffer
zone $m_1\ll r_1\ll b$.  The largest errors come from inaccuracies in the coordinate
transformation~(\ref{finaltrans}).  Terms of the form $(m^2/b^2)(r_1/b)^p$ for integer $p\geq 1$
have been ignored in Eq.~(\ref{finaltrans}).  This leads to errors $\dg_{\mu\nu}\sim\eps^4$
for $r_1\lesssim r_1^{in}$.

The match between the internal and post-Newtonian metrics at $r_1=r_1^{in}$ is not perfect; there
are discontinuities $[g_{\mu\nu}]$ in the metric components on that 3-surface.  A term of
the form $m_1^3/r_1^3$ in the internal metric component $g_{00}$ [as given in Eq.~(\ref{gPKI})]
is not matched in the post-Newtonian metric component $g_{00}$ in Eq.~(\ref{corotmetnear});
as a result, $[g_{00}]\sim\eps^3$ at $r_1=r_1^{in}$.
Similarly, a term of the form $m_1^2/r_1^2$ is not matched
in $g_{ij}$, so $[g_{ij}]\sim\eps^2$.  Lastly, a term of the form $m_1^2/R^2$ in the
internal-coordinate metric component $g_{00}$ in Eq.~(\ref{intmet}) gives via a (partial)
Lorentz boost an unmatched term of the form $\eps m_1^2/r_1^2$ in the internal-metric component
$g_{0i}$ in corotating coordinates [given in Eq.~(\ref{gPKI})]; so $[g_{0i}]\sim\eps^3$.

The internal metric~(\ref{gPKI}) contains terms of the form $(m_1/r_1)^p(m_2 r_1^2/b^3)$,
$p\geq 1$, in the buffer zone.  These terms represent the first black hole's multipole
moments and the nonlinear interaction of internal and external gravitational fields.  They are
of $O(\eps^{p+4})$, $p\geq 1$, in the buffer zone and have not been matched to the post-Newtonian
near-zone metric~(\ref{corotmetnear}).  At the level of accuracy achieved in this paper, the
metric~(\ref{corotmetnear}) need not be modified to include the near-zone
gravitational effects of the black holes' deformation.

\subsection{Metric near the second black hole}

An identical procedure can now be followed to obtain the metric in corotating
post-Newtonian coordinates in region II.  However, it is not
necessary to repeat all the steps.  This metric can simply be gotten as follows:
exchange $m_1 \leftrightarrow m_2$ in the internal metric~(\ref{intmet}) and in the
transformation~(\ref{finaltrans}); take $x \to -x$ and $y \to -y$ in Eq.~(\ref{finaltrans});
then transform.  In other words, the metric components in region II
$\left[\textrm{denoted }g^{I\!I}_{\mu\nu}(t,x,y,z)\right]$ are related to those
in region I $\left[g^I_{\mu\nu}(t,x,y,z)\right]$ by
$g^{I\!I}_{\mu\nu}(t,x,y,z)=(-1)^p g^{I}_{\mu\nu}(t,-x,-y,z)
({\rm with\ }m_1\leftrightarrow m_2)$, where $p$ is the number of
the indices $\mu$ and $\nu$ that are equal to 1 or 2.

Define $\bar{P}_{\alpha\beta}$ to be
$P_{\alpha\beta}$ with $m_1$ and $m_2$ exchanged, and similarly $\bar{K}^{\sigma}_{\rho}$
to be $K^{\sigma}_{\rho}$ with $m_1\leftrightarrow m_2$.  Then the metric in region II
is given in corotating coordinates $(t,x,y,z)$ by
\begin{equation}
	g_{\mu\nu}(x,y,z) = (-1)^p\sum_{\alpha,\sigma=0}^{3} \bar{P}_{\alpha\sigma}(-x,-y,z)
		\bar{K}^{\alpha}_{\mu}(-x,-y,z)\bar{K}^{\sigma}_{\nu}(-x,-y,z)
\label{gPKII}
\end{equation}
where $p$ is, as above, the number of the indices $\mu$ and $\nu$ that are equal
to 1 or 2.  Again, the metric components are explicitly independent of time $t$.
This metric is valid up to the second black hole's apparent horizon and matches
(to a finite order) the post-Newtonian metric~(\ref{corotmetnear}) at $r_2=r_2^{in}$.  Error
analysis for this metric is identical to the analysis above for the metric in region I.

\section{Results and discussion}
\label{results}

The result of this calculation is an approximate solution to
Einstein's equations representing two widely-separated
non-rotating black holes in a circular orbit.  The metric has
been expressed in a single set of coordinates valid up to the
black holes' apparent horizons; the coordinate system chosen is corotating
coordinates $(t,x,y,z)$.  In these coordinates, the metric components are explicitly
independent of time $t$.  The metric is specified in region I by
Eq.~(\ref{gPKI}), in region II by Eq.~(\ref{gPKII}), in region III by Eq.~(\ref{corotmetnear}),
and in region IV by Eq.~(\ref{corotmetfar}).  At the boundaries $r_1=r_1^{in}$, $r_2=r_2^{in}$,
and $r=r^{out}$ of these regions, there are discontinuities in the metric components that
result from truncation of perturbative expansions and
finite-order matching.  The magnitudes of these discontinuities can be reduced by taking this
calculation to higher orders.

The full 4-metric is summarized below:
\begin{eqnarray}
g_{00} &=& \cases{\sum_{\mu,\nu=0}^{3} P_{\mu\nu}(x,y,z)K^{\mu}_{0}(x,y,z)K^{\nu}_{0}(x,y,z),
	& in region I;\cr
	\sum_{\mu,\nu=0}^{3} \bar{P}_{\mu\nu}(-x,-y,z)
		\bar{K}^{\mu}_{0}(-x,-y,z)\bar{K}^{\nu}_{0}(-x,-y,z),
	& in region II;\cr
	-1 + {2m_1 \over r_1} + {2m_2 \over r_2}-2\left({m_1 \over r_1} + {m_2 \over r_2}\right)^2
	+ {3\mu\over b}\left({m_2 \over r_1} + {m_1 \over r_2}\right)
	- {\mu\over b}\left({m_2 \over r_1^3} + {m_1 \over r_2^3}\right)y^2 &\cr
\qquad\mbox{}-2\mu\eps^2\left({1\over r_1}+{1\over r_2}\right)
	- 7\mu\eps^2\left({1\over r_1}-{1\over r_2}\right){x\over b}
	+\w^2\left(1 + {2m_1 \over r_1}+{2m_2 \over r_2} \right)(x^2 + y^2),
	& in region III;\cr
	-1 + {2m \over r}\left(1-{\mu\over 2b}\right)-{2m^2 \over r^2}+A
		+ {2\eps^2\over b}(x\cos\w r-y\sin\w r)B &\cr
\qquad\mbox{}-2\eps(x\sin\w r+y\cos\w r)D+\w^2(x^2+y^2)E &\cr
\qquad\mbox{}+{\eps^4\over b^2}(x\cos\w r-y\sin\w r)^2 N
		+\eps^2(x\sin\w r+y\cos\w r)^2S &\cr
\qquad\mbox{}-{12\mu\eps^5\over r^2 b^2}{\delta m\over m}(x\cos\w r-y\sin\w r)^2
		(x\sin\w r+y\cos\w r),
	& in region IV.\cr}\label{g00final}\\
g_{0x} &=& \cases{\sum_{\mu,\nu=0}^{3} P_{\mu\nu}(x,y,z)K^{\mu}_{0}(x,y,z)K^{\nu}_{1}(x,y,z),
	& in region I;\cr
	-\sum_{\mu,\nu=0}^{3} \bar{P}_{\mu\nu}(-x,-y,z)\bar{K}^{\mu}_{0}(-x,-y,z)
		\bar{K}^{\nu}_{1}(-x,-y,z),
	& in region II;\cr
     	-\w y\left(1 + {2m_1 \over r_1}+{2m_2 \over r_2}\right),
	& in region III;\cr
	\eps B\sin\w r + bD\cos\w r - \w Ey+{\eps^3\over b}(x\cos\w r-y\sin\w r)N\sin\w r &\cr
\qquad\mbox{}-\eps b(x\sin\w r+y\cos\w r)S\cos\w r &\cr
\qquad\mbox{}+{6\mu\eps^4\over r^2 b}{\delta m\over m}(x\cos\w r-y\sin\w r)(x\cos 2\w r-y\sin 2\w r),
	& in region IV.\cr}\label{g0xfinal}\\
g_{0y} &=& \cases{\sum_{\mu,\nu=0}^{3} P_{\mu\nu}(x,y,z)K^{\mu}_{0}(x,y,z)K^{\nu}_{2}(x,y,z),
	& in region I;\cr
	-\sum_{\mu,\nu=0}^{3} \bar{P}_{\mu\nu}(-x,-y,z)\bar{K}^{\mu}_{0}(-x,-y,z)
		\bar{K}^{\nu}_{2}(-x,-y,z),
	& in region II;\cr
     	\w x\left(1 + {2m_1 \over r_1}+{2m_2 \over r_2}\right)
		-4\mu\eps\left({1\over r_1}-{1\over r_2}\right),
	& in region III;\cr
	\eps B\cos\w r - bD\sin\w r + \w Ex+{\eps^3\over b}(x\cos\w r-y\sin\w r)N\cos\w r &\cr
\qquad\mbox{}+\eps b(x\sin\w r+y\cos\w r)S\sin\w r &\cr
\qquad\mbox{}-{6\mu\eps^4\over r^2 b}{\delta m\over m}(x\cos\w r-y\sin\w r)(x\sin 2\w r+y\cos 2\w r),
	& in region IV.\cr}\label{g0yfinal}\\
g_{0z} &=& \cases{\sum_{\mu,\nu=0}^{3} P_{\mu\nu}(x,y,z)K^{\mu}_{0}(x,y,z)K^{\nu}_{3}(x,y,z),
	& in region I;\cr
	\sum_{\mu,\nu=0}^{3} \bar{P}_{\mu\nu}(-x,-y,z)\bar{K}^{\mu}_{0}(-x,-y,z)
		\bar{K}^{\nu}_{3}(-x,-y,z),
	& in region II;\cr
	0,
	& in regions III and IV.\cr}\label{g0zfinal}\\
g_{xx} &=& \cases{\sum_{\mu,\nu=0}^{3} P_{\mu\nu}(x,y,z)K^{\mu}_{1}(x,y,z)K^{\nu}_{1}(x,y,z),
	& in region I;\cr
	\sum_{\mu,\nu=0}^{3} \bar{P}_{\mu\nu}(-x,-y,z)\bar{K}^{\mu}_{1}(-x,-y,z)
		\bar{K}^{\nu}_{1}(-x,-y,z),
	& in region II;\cr
	1 + {2m_1 \over r_1} + {2m_2 \over r_2},
	& in region III;\cr
	E+{m^2 x^2\over r^4}+\eps^2 N\sin^2\w r+b^2 S\cos^2\w r+{6\mu\eps^3\over r^2}
		{\delta m\over m}(x\cos\w r-y\sin\w r)\sin 2\w r,
	& in region IV.\cr}\label{gxxfinal}\\
g_{yy} &=& \cases{\sum_{\mu,\nu=0}^{3} P_{\mu\nu}(x,y,z)K^{\mu}_{2}(x,y,z)K^{\nu}_{2}(x,y,z),
	& in region I;\cr
	\sum_{\mu,\nu=0}^{3} \bar{P}_{\mu\nu}(-x,-y,z)\bar{K}^{\mu}_{2}(-x,-y,z)
		\bar{K}^{\nu}_{2}(-x,-y,z),
	& in region II;\cr
	1 + {2m_1 \over r_1} + {2m_2 \over r_2},
	& in region III;\cr
	E+{m^2 y^2\over r^4}+\eps^2 N\cos^2\w r+b^2 S\sin^2\w r-{6\mu\eps^3\over r^2}
		{\delta m\over m}(x\cos\w r-y\sin\w r)\sin 2\w r,
	& in region IV.\cr}\label{gyyfinal}\\
g_{zz} &=& \cases{\sum_{\mu,\nu=0}^{3} P_{\mu\nu}(x,y,z)K^{\mu}_{3}(x,y,z)K^{\nu}_{3}(x,y,z),
	& in region I;\cr
	\sum_{\mu,\nu=0}^{3} \bar{P}_{\mu\nu}(-x,-y,z)\bar{K}^{\mu}_{3}(-x,-y,z)
		\bar{K}^{\nu}_{3}(-x,-y,z),
	& in region II;\cr
	1 + {2m_1 \over r_1} + {2m_2 \over r_2},
	& in region III;\cr
	E+{m^2 z^2\over r^4},
	& in region IV.\cr}\label{gzzfinal}\\
g_{xy} &=& \cases{\sum_{\mu,\nu=0}^{3} P_{\mu\nu}(x,y,z)K^{\mu}_{1}(x,y,z)K^{\nu}_{2}(x,y,z),
	& in region I;\cr
	\sum_{\mu,\nu=0}^{3} \bar{P}_{\mu\nu}(-x,-y,z)\bar{K}^{\mu}_{1}(-x,-y,z)
		\bar{K}^{\nu}_{2}(-x,-y,z),
	& in region II;\cr
	0,
	& in region III;\cr
	{m^2\over r^4}xy+{1\over 2}(\eps^2 N - b^2 S)\sin 2\w r &\cr
\qquad\mbox{}+{6\mu\eps^3\over r^2}{\delta m\over m}(x\cos\w r-y\sin\w r)\cos 2\w r,
	& in region IV.\cr}\label{gxyfinal}\\
g_{xz} &=& \cases{\sum_{\mu,\nu=0}^{3} P_{\mu\nu}(x,y,z)K^{\mu}_{1}(x,y,z)K^{\nu}_{3}(x,y,z),
	& in region I;\cr
	-\sum_{\mu,\nu=0}^{3} \bar{P}_{\mu\nu}(-x,-y,z)\bar{K}^{\mu}_{1}(-x,-y,z)
		\bar{K}^{\nu}_{3}(-x,-y,z),
	& in region II;\cr
	0,
	& in region III;\cr
	{m^2\over r^4}xz,
	& in region IV.\cr}\label{gxzfinal}\\
g_{yz} &=& \cases{\sum_{\mu,\nu=0}^{3} P_{\mu\nu}(x,y,z)K^{\mu}_{2}(x,y,z)K^{\nu}_{3}(x,y,z),
	& in region I;\cr
	-\sum_{\mu,\nu=0}^{3} \bar{P}_{\mu\nu}(-x,-y,z)\bar{K}^{\mu}_{2}(-x,-y,z)
		\bar{K}^{\nu}_{3}(-x,-y,z),
	& in region II;\cr
	0,
	& in region III;\cr
	{m^2\over r^4}yz,
	& in region IV.\cr}\label{gyzfinal}
\end{eqnarray}
In the expressions above, $m=m_1+m_2$, $\mu=m_1 m_2/m$, $\delta m=m_1-m_2$, $\eps=(m/b)^{1/2}$,
$\w=(m/b^3)^{1/2}$, $r_1=[(x-m_2 b/m)^2+y^2+z^2]^{1/2}$, and
$r_2=[(x+m_1 b/m)^2+y^2+z^2]^{1/2}$.  Region I is specified by $r_1<(m_1 b)^{1/2}$ and
region II by $r_2<(m_2 b)^{1/2}$ (but these regions do not extend
inside the black holes' apparent horizons).  Region III is specified by $r_1>(m_1 b)^{1/2}$,
$r_2>(m_2 b)^{1/2}$, and $r<b/2\eps$; and region IV by $r>b/2\eps$.  The functions $A$, $B$,
$D$, $E$, $N$, and $S$ are defined in Eq.~(\ref{ABDENS}).
The functions $P_{\alpha\beta}$ and $K^{\sigma}_{\rho}$
are defined in Eqs.~(\ref{P}) and (\ref{K}).  The functions $\bar{P}_{\alpha\beta}$ and
$\bar{K}^{\sigma}_{\rho}$ are obtained from $P_{\alpha\beta}$ and $K^{\sigma}_{\rho}$
respectively by exchanging $m_1$ and $m_2$.

The errors and discontinuities in the metric components are summarized in Table~\ref{errors}.
The discontinuities should
be smoothed out before initial data are extracted from the metric.  In addition,
initial data taken should be relaxed numerically to approach more closely an exact
solution of the constraint equations.  It is expected that
higher-order versions of this calculation will differ by smaller
amounts from an exact solution of the Einstein equations.

\renewcommand{\arraystretch}{1.6}
\newlength{\templength}
\setlength{\templength}{\tabcolsep}
\setlength{\tabcolsep}{0.28cm}
\begin{center}
\begin{tabular}{|l|c|c|c|c|c|c|c|c|} \hline
 & Region I or II & disc. at & \multicolumn{3}{c|}{Region III} & &
	\multicolumn{2}{c|}{Region IV} \\ \cline{2-2} \cline{4-6} \cline{8-9}
 & $r_1\lesssim r_1^{in}$ & $r_1=r_1^{in}$ & $r_1\gtrsim r_1^{in}$ & $r_1\sim b$ & & disc. at & & \\
 & or & or & or & and & $r\lesssim r^{out}$ & $r=r^{out}$ & $r\gtrsim r^{out}$ & $r\sim b/\epsilon^2$ \\
 & $r_2\lesssim r_2^{in}$ & $r_2=r_2^{in}$ & $r_2\gtrsim r_2^{in}$ & $r_2\sim b$ & & & & \\ \hline
$g_{00}$ & 4 & 3 & 3 & 6 & 5 & 5 & 7 & 8 \\ \hline
$g_{0i}$ & 4 & 3 & 3 & 5 & 5 & 5 & 7 & 8 \\ \hline
$g_{ij}$ & 4 & 2 & 2 & 4 & 5 & 5 & 7 & 8 \\ \hline
\end{tabular}
\end{center}
\renewcommand{\arraystretch}{1}
\setlength{\tabcolsep}{\templength}
\begin{table}[h]
\caption{Errors and discontinuities in the metric components in corotating coordinates.  Numbers
	denote orders in $\eps=(m/b)^{1/2}$; e.g., 4 denotes $O(\eps^4)$.  The last two columns
	contain normalized errors.}
\label{errors}
\end{table}

\section*{Acknowledgments}

I would like to thank Kip Thorne for suggesting this research
project and for his guidance during its completion.  This research was
supported in part by NASA grant NAG5-6840 and by NSF grant AST-9731698.

\end{document}